\documentclass[reprint, longbibliography, amsmath, amssymb, aps, prx]{revtex4-2}
\usepackage{array}
\usepackage{autobreak}
\usepackage{graphicx}%
\usepackage{dcolumn}%
\usepackage{tabularx}
\usepackage{booktabs}
\usepackage{bm}%
\usepackage{float}
\usepackage{braket}
\usepackage{multirow}
\usepackage{newtxtext}
\usepackage{xcolor}
\usepackage[normalem]{ulem}
\usepackage{here}
\usepackage{comment}

\begin{document}
\title{Impact of electron--spin coupling on exchange coupling parameters: a nonperturbative approach}
\author{Tomonori Tanaka}
 \email{tanaka.t.da74@m.isct.ac.jp}
\author{Yoshihiro Gohda}
 \email{gohda@mct.isct.ac.jp}
\affiliation{Department of Materials Science and Engineering, Institute of Science Tokyo, Yokohama 226-8501, Japan}
\date{\today}%
\begin{abstract}
Exchange coupling parameters $J_{ij}$ in the Heisenberg model are crucial for describing magnetic behavior at the atomic level.
In magnetic materials, spin fluctuations can be accompanied by a self-consistent electronic response, i.e., electron--spin coupling in the sense of electronic feedback to finite spin rotations.
However, the quantitative importance of this coupling in extracting reliable $J_{ij}$ has not been fully clarified.
Here, using fully self-consistent, nonperturbative evaluations, we show that finite-angle spin rotations induce such electronic feedback and quantify how strongly it renormalizes the extracted $J_{ij}$.
We examine systems of both fundamental and practical interest, including perovskite SrMnO$_3$, Nd-based permanent-magnet compounds (Nd$_2$Fe$_{14}$B and Nd$_2$Co$_{14}$B), and elemental 3$d$ transition metals.
The nonperturbative approach yields exchange couplings that remain consistent over a wide range of rotation angles.
Moreover, when the sampled configurations represent the magnetic disorder near the transition, spin models parameterized in this way give reasonable agreement with experimental magnetic phase-transition temperatures.
Overall, our results demonstrate the quantitative impact of electron--spin coupling on the exchange coupling parameters $J_{ij}$ and provide a nonperturbative route to effective $J_{ij}$ for a specified regime of magnetic disorder.
\end{abstract}
\maketitle
\clearpage
\section{Introduction}
Understanding magnetic interactions at the atomic level is increasingly important for quantitative modeling and magnetic-materials design.
It is also relevant to spintronics, where magnetic textures, their dynamics, and their coupling to electronic transport often play key roles in device functionalities~\cite{Dey2021-qx}.
Since many key phenomena---magnetic phase transitions, thermally activated disorder, and dynamical spin textures---occur at finite temperature and involve large-scale collective fluctuations, it is often impractical to treat them directly within first-principles electronic-structure calculations~\cite{Eriksson2017-lp}.
A common and effective strategy is therefore to coarse-grain the problem and map it onto a classical spin Hamiltonian, which retains the orientations of local magnetic moments as the essential degrees of freedom while integrating out electronic and longitudinal fluctuations, such that the energy of a spin configuration is represented by effective interaction parameters~\cite{Drautz2004-id}.

The central challenge then becomes how to determine effective interaction parameters from first principles in a manner that remains valid beyond a narrow neighborhood of a reference magnetic state.
When finite spin rotations lead to noticeable changes in the underlying electronic structure, they are accompanied by a self-consistent electronic response---such as charge and magnetization redistribution and changes in orbital occupations---thereby coupling spin orientations to the electronic structure.
Here, we use the term electron--spin coupling to denote this self-consistent electronic response to spin rotations and its energetic feedback on the effective spin-model parameters.
This feedback can be effectively folded into the extracted parameters under finite-angle spin fluctuations~\cite{Bottcher2012-bl, Szilva2013-of, Kvashnin2016-bl, Szilva2017-sw, Cardias2017-kh, Chimata2017-du} and magnetic disorder, potentially impacting quantitative predictions such as magnetic phase-transition temperatures.

A widely used route to first-principles interaction parameters is provided by perturbative mappings based on the magnetic force theorem (MFT)~\cite{Oguchi1983-vi, Liechtenstein1987-ev}, most notably the Liechtenstein method.
Within the infinitesimal-rotation limit around a chosen reference state, this framework offers an efficient and physically transparent way to extract exchange interactions~\cite{Katsnelson2004-pw, Solovyev1998-jx}, has been further developed and extended in various directions~\cite{Szilva2013-of, Kvashnin2015-ky, Kvashnin2016-bl, Szilva2017-sw, Yoon2018-ik, Terasawa2019-mb, Nomoto2020-ru, He2021-dw, Solovyev2021-bz}, and has been applied to a broad range of magnetic materials~\cite{Mankovsky2022-xy, Szilva2023-ta}.
In this formulation, the mapping is carried out by evaluating the band-energy response to infinitesimal spin rotations while keeping the charge density and the magnitude of the magnetization fixed, and the accompanying self-consistent changes in these quantities enter only at higher order.
The quantitative consequences of this approximation---in particular, when and how strongly electron--spin coupling renormalizes the extracted exchange parameters under finite-angle spin rotations---have not been systematically established outside a limited set of benchmark elemental magnets~\cite{Bottcher2012-bl, Szilva2013-of, Kvashnin2016-bl, Szilva2017-sw, Cardias2017-kh, Chimata2017-du}.
Even within an MFT-based framework, analyses of itinerant metallic magnetic materials have long emphasized the sensitivity of extracted interactions to electronic-structure details, including orbital occupations and Fermi-surface properties~\cite{Szilva2013-of, Kvashnin2016-bl, Szilva2017-sw}.
A fully self-consistent, nonperturbative treatment~\cite{Antal2008-nt, Drautz2004-id, Mendive-Tapia2022-wq} is therefore valuable for quantifying how such electronic-structure sensitivities translate into effective interaction parameters beyond the strict infinitesimal-rotation regime.

Previous first-principles studies have reported that conventional MFT-based mappings fail to reproduce the total-energy variation under small but finite spin rotations in type--G antiferromagnetic SrMnO$_3$~\cite{Zhu2020-ik} and the experimentally observed increase in $T_{\rm C}$ upon Co substitution in Nd$_2$(Fe,Co)$_{14}$B~\cite{Harashima2021-iz}.
Although the origin of such discrepancies can be system dependent, these observations suggest that electron--spin coupling may renormalize effective spin-model parameters beyond what is captured in infinitesimal-rotation-based treatments.

In this study, we demonstrate how electron--spin coupling affects the exchange parameters $J_{ij}$ in a Heisenberg-model description by using a fully self-consistent, nonperturbative mapping, and quantify how strongly $J_{ij}$ is renormalized under finite-angle spin fluctuations and magnetic disorder and how this impacts finite-temperature magnetism.
After presenting the method and its relation to other nonperturbative approaches (Sec.~\ref{theory}) and the computational conditions (Sec.~\ref{conditions}), we show that the discrepancy in SrMnO$_3$ originates from spin-rotation-induced changes in orbital occupations (Sec.~\ref{srmno3}), that the method captures the Co-substitution trend in $T_{\rm C}$ of the Nd-based compounds (Sec.~\ref{2-14-1}), and that $J_{ij}$ in elemental 3$d$ magnets depends on the rotation angle much more strongly than is commonly assumed (Sec.~\ref{3d}).
A conceptual discussion in the language of MFT is given in Sec.~\ref{discussion}, followed by a summary in Sec.~\ref{summary}.

\section{Methods}
\label{theory}
\subsection{Nonperturbative calculation of $J_{ij}$ with supercells}
The validity of an effective spin model depends on the symmetry of the system and the required energy accuracy.
We use the classical Heisenberg model consisting of the lowest-order pairwise interactions throughout this study as
\begin{align}
E(\{\hat{{\bm e}}_i\}) = E_0 -\sum_{i\neq j} J_{ij} \hat{{\bm e}}_{i} \cdot \hat{\bm e}_{j},
\label{heisenberg}
\end{align}
where $E_0$ is the reference energy, $J_{ij}$ is the isotropic exchange coupling parameter between sites $i$ and $j$, and $\hat{\bm e}_i$ is a unit vector representing the orientation of the local magnetic moment on site $i$.
This simplified model is valid when the influence of the anisotropy of the system on the physical quantities of interest is small.
The model parameters $J_{ij}$ and $E_0$ can be determined by minimizing the mean squared error~\cite{Antal2008-nt} defined as
\begin{align}
\Delta E^2 = \frac{1}{N_{\rm data}}\sum_{n=1}^{N_{\rm data}} \left( E_{\rm DFT}^{(n)} - E_{\rm model}^{(n)}\right)^2,
\label{least-squares}
\end{align}
where $E_{\rm DFT}^{(n)}$ is the energy obtained from density functional theory (DFT) calculations for the $n$-th magnetic configuration, $E_{\rm model}^{(n)}$ is the energy related to the spin model, and $N_{\rm data}$ is the total number of magnetic configurations.
In this study, we used the supercell method to model magnetic configurations.
Therefore, we hereafter refer to this procedure as the (SC)$^2$ method (Self-Consistent SuperCell method) for convenience.
In the practical calculations, we used the symmetry of the target systems to identify independent $J_{ij}$ with the Spglib library~\cite{Togo2024-vg}.

The model parameters in Eq.~(\ref{heisenberg}) depend on how the magnetic configuration dataset is sampled.
Our focus is on $J_{ij}$ near stable or metastable magnetic configurations to compare them with those obtained from MFT-based methods.
Therefore, we slightly tilted the direction of the magnetic moment on each atom from a (meta)stable magnetic configuration.
Specifically, we used uniformly distributed random numbers on a spherical cap in the range of $0^{\circ} \leqq \theta \leqq \theta_{\rm max}$ by using inverse transform sampling, while azimuthal angles were unrestricted.
Magnetic configurations were sampled until the parameters converged; typically, we used three times the number of independent $J_{ij}$ parameters.
The above sampling procedure leads to the dependence of $J_{ij}$ on $\theta_{\rm max}$.
This dependence can arise from electron--spin coupling induced by magnetic disorder, i.e., spin-disorder-driven changes in the electronic structure that feed back on the effective $J_{ij}$.
Consequently, the $\theta_{\rm max}$ dependence of $J_{ij}$ provides insight into how such electron--spin coupling renormalizes $J_{ij}$ in disordered environments and, in turn, affects the resulting transition temperature.
This will be revisited in the discussion of $T_{\rm C}$ in Nd$_2$Fe$_{14}$B and Nd$_2$Co$_{14}$B (Sec.~\ref{2-14-1}).

If one wishes to replace the cutoff angle $\theta_{\rm max}$ by a physically motivated parameter, a more refined, temperature-controlled sampling scheme can be used.
For example, one may sample spin configurations according to the spin-direction distribution derived from the mean-field Heisenberg model~\cite{Mendive-Tapia2022-wq}:
\begin{align}
	P_i(\hat{\bm e}_i)=\frac{3m/\tau}{4\pi\sinh(3m/\tau)}
\exp\!\left(\frac{3\,{\bm m}\!\cdot\!\hat{\bm e}_i}{\tau}\right),
\label{spin_dist}
\end{align}
where $P_i(\hat{\bm e}_i)$ is the normalized probability density on the unit sphere for the spin orientation at site $i$, ${\bm m} = m\hat{\bm e}$ is the site-independent thermal average of the spin, and $\tau = T/T_{\rm C}^{\rm MFA}$ is a reduced temperature (with $T$ the temperature and $T_{\rm C}^{\rm MFA}$ the mean-field critical temperature).
Introducing $\tau$ enables sampling from a temperature-controlled distribution. In the small-angle regime, the two schemes, when matched in $\langle \hat{\bm e}_i \cdot \hat{\bm e}_j \rangle$, yield consistent $J_{ij}$ for bcc Fe and fcc Ni (tested at $\theta_{\rm max}=20^\circ$ and the corresponding $\tau$; Sec.~S11~\cite{SI}), so that the $\theta_{\rm max}$ dependence of $J_{ij}$ discussed below for small angles does not depend on the details of the sampling scheme.
As a reference, $\theta_{\rm max}=10^\circ$, $20^\circ$, and $30^\circ$ correspond to $\langle \hat{\bm e}_i \cdot \hat{\bm e}_j \rangle = 0.98$, $0.94$, and $0.87$, whereas the $\tau$ values used in Figs.~\ref{tc_2-14-1} and \ref{tc_elements} correspond to $\langle \hat{\bm e}_i \cdot \hat{\bm e}_j \rangle = 0.5$ and $0.65$.
Within the mean-field theory, the spin-direction distribution becomes uniform at $T_{\rm C}^{\rm MFA}$, corresponding to a completely disordered limit.
Therefore, if one aims to sample spin configurations representative of the true magnetic phase transition temperature, one should choose $\tau$ (or the corresponding $m$) such that a finite degree of short-range order is retained.
In the present work, $\tau$ is treated as an input that specifies the regime of magnetic disorder for which the effective $J_{ij}$ are constructed, not a quantity determined within the method itself.
To obtain more quantitative estimates of the transition temperatures, we also used $\tau$ values chosen such that the nearest-neighbor spin--spin correlation function $\langle \hat{\bm e}_i \cdot \hat{\bm e}_j \rangle$ matches values reported near $T_{\rm C}$ in previous studies based on Monte Carlo simulations and dynamic spin-fluctuation theory~\cite{Melnikov2019-es, Walsh2022-ci}.
Specifically, for the Nd-based permanent-magnet compounds and bcc Fe we used $\langle \hat{\bm e}_i \cdot \hat{\bm e}_j \rangle = 0.5$ ($\tau = 0.6261$), whereas for fcc Co and fcc Ni we used $\langle \hat{\bm e}_i \cdot \hat{\bm e}_j \rangle = 0.65$ ($\tau = 0.4688$).
See also Sec.~S11 of the Supplementary Information~\cite{SI} for a detailed description of the sampling strategy constructed from the mean-field Heisenberg model.

The parameters $J_{ij}$ in the (SC)$^2$ method involve a size effect similar to that in the direct method for phonon calculations; the parameters are not values between individual sites but between the periodic images of sites $i$ and $j$.
This size effect decreases as the supercell size increases.
Therefore, we carefully check the convergence of $J_{ij}$ with respect to the supercell size.
Additionally, we distinguish the (SC)$^2$ method from approaches that rely on energy differences between completely different magnetic states, such as ferromagnetic and antiferromagnetic states.
The latter method assumes that the magnetic coupling parameters remain constant regardless of the magnetic configuration. 
However, in many metallic systems this assumption does not hold.
For example, in bcc Fe the magnetic moment differs markedly between the ferromagnetic state ($\sim$2.2 $\mu_{\rm B}$/atom) and the type--I antiferromagnetic state (1.3 $\mu_{\rm B}$/atom)~\cite{Herper1999-nm}, rendering any single Heisenberg model fitted across these states unreliable.
On the other hand, the (SC)$^2$ method determines $J_{ij}$ by sampling relevant magnetic configurations while verifying whether the Heisenberg model can accurately reproduce the energy dataset.

Although we focus on magnetic configurations near the ground state, the (SC)$^2$ method is also applicable to the Curie--Weiss paramagnetic state by setting $\theta_{\rm max} = 180^{\circ}$.
Such paramagnetic states have also been studied in the context of the Korringa--Kohn--Rostoker (KKR) Green's function method and the coherent potential approximation~\cite{Oguchi1983-vi}.
The results of these two methods for bcc Fe are presented in Sec.~S9 of the Supplementary Information~\cite{SI}. 

Related approaches to the calculation of $J_{ij}$ based on self-consistent calculations have also been reported in the literature.
For example, spin configurations can be sampled using spin-spiral approaches~\cite{Sandratskii1998-ln, Jacobsson2022-nu}.
In addition, methods known as spin-cluster expansion have been proposed~\cite{Drautz2004-id}, in which the spin model is not restricted to the Heisenberg form but is instead assumed to have a general functional form, with the corresponding parameters determined by fitting to total energies.
Comparisons with, and the relationship to, these approaches are discussed in Sec.~\ref{spin-spiral} and Sec.~\ref{sce}.

\subsection{Comparison with the spin-spiral method}
\label{spin-spiral}
We briefly review the differences between the (SC)$^2$ method and the spin-spiral method, both of which are nonperturbative approaches.
The spin-spiral method, based on the generalized Bloch's theorem~\cite{Sandratskii1998-ln}, is another procedure for sampling magnetic configurations~\cite{Jacobsson2022-nu}. Subsequently, exchange coupling parameters $J_{ij}$ are evaluated to reproduce the energies of sampled spin-spiral states.
A major advantage of this method is that it requires only a minimal magnetic unit cell, thus keeping computational costs relatively low.
However, there are limitations not present in the (SC)$^2$ method.
For example, the spin-spiral method does not explicitly capture the disorder arising from the simultaneous excitation of multiple spin-wave modes and the associated electron--spin coupling, which becomes unavoidable at finite temperatures.
Moreover, spin-orbit coupling cannot be incorporated alongside the spin-spiral method because it breaks the translational symmetry assumed in the generalized Bloch's theorem.

The (SC)$^2$ method is also extensible.
First, it is not limited to evaluating the isotropic lowest-order cluster interaction $J_{ij}$.
For instance, Antal and co-workers have evaluated tensor-formed exchange coupling parameters, fourth-order interaction parameters, and on-site anisotropy parameters for a Cr trimer on an Au(111) surface while considering the spin-orbit coupling~\cite{Antal2008-nt}.
Second, the effect of lattice vibrations on exchange coupling parameters can also be calculated straightforwardly, as demonstrated in Ref.~\cite{Heine2021-qe}.
The direct modeling of atomic displacements due to lattice vibrations in a supercell is a feature not available in the spin-spiral method with a minimal cell.
The current limitation of the (SC)$^2$ method compared to other methods is its higher computational cost and the difficulty in systematically improving spin models.
However, we expect these issues can be overcome in the future by combining the spin-cluster expansion~\cite{Drautz2004-id} with the torque method that utilizes constraining magnetic fields~\cite{Jacobsson2022-nu}.

\subsection{Relation to spin-cluster expansion}
\label{sce}
The spin-cluster expansion method~\cite{Drautz2004-id} provides a general framework for constructing classical spin models.
In this method, the energy of a spin model is expressed as
\begin{align}
E(\{\hat{\bm e}_i\}) = E_0 + \sum_{\alpha}\sum_{\bm l}\sum_{\bm m} J_{\alpha \bm l \bm m}\,\bm{\Phi}_{\alpha \bm l \bm m}(\{\hat{\bm e}_i\}),
\label{eq:sce}
\end{align}
where $E_0$ is a reference energy, $\alpha$ labels a cluster (a set of sites), the multi-indices ${\bm l}=(l_1,\ldots,l_{n_\alpha})$ and ${\bm m}=(m_1,\ldots,m_{n_\alpha})$ collect the spherical-harmonic degrees and orders for the $n_\alpha$ sites in cluster $\alpha$, and $J_{\alpha \bm l \bm m}$ are the corresponding cluster-interaction coefficients.
The cluster basis functions are defined as
\begin{align}
\bm{\Phi}_{\alpha \bm l \bm m}(\{\hat{\bm e}_i\}) = \prod_{i\in\alpha} y_{l_i m_i}(\hat{\bm e}_i),
\end{align}
where $y_{l m}$ are the spherical harmonics with indices ($l, m$).
Note that the $y_{l m}$ are defined in terms of the conventional spherical harmonics $Y_{l m}$ as
\begin{align}
y_{l m} = \sqrt{4\pi} Y_{l m}.
\end{align}

In practical applications, the number of clusters must be finite.
In this case, the parameter set \{$J_{\alpha \bm{l} \bm{m}}$\} depends on the sampled magnetic configurations and the selected clusters.
The dependence on the selection of clusters diminishes as the number of considered clusters increases; however, this comes at the cost of a more complex model.
Therefore, as long as the sampled dataset (a set of magnetic configurations and corresponding energies) obtained from DFT is sufficiently reproduced, it would be better to construct a model that effectively renormalizes higher-order cluster interactions using only a small number of lower-order clusters.
This discussion applies more generally to series expansions and the corresponding function fitting.

The Heisenberg model in Eq.~(\ref{heisenberg}) is a special case of the spin-cluster expansion model.
The inner product between two spins ($\hat{{\bm e}}_{i}\cdot\hat{{\bm e}}_{j}$) in the Heisenberg model can be incorporated into the spin-cluster expansion as follows:
\begin{align}
\hat{{\bm e}}_{i} \cdot \hat{{\bm e}}_{j} = \frac{1}{3} \sum_{m = -1}^{1} y_{1 m}^{*}(\hat{\bm e}_i) y_{1 m}(\hat{\bm e}_j).
\end{align}
This relationship is straightforwardly derived from the spherical harmonics addition theorem;
\begin{align}
P_{l}(\hat{{\bm e}}_{i}\cdot\hat{{\bm e}}_{j}) = \frac{1}{2l + 1}\sum_{m = -l}^{l} y_{l m}^{*}(\hat{\bm e}_i) y_{l m}(\hat{\bm e}_j),
\end{align}
where $P_{l}$ denotes the $l$-th Legendre polynomial.
Although the Heisenberg model corresponds to the lowest-order cluster interactions within the spin-cluster expansion framework, higher-order interactions are renormalized into the $J_{ij}$ parameters in Eq.~(\ref{heisenberg}) through the fitting procedure in Eq.~(\ref{least-squares}).
Therefore, the $J_{ij}$ parameters naturally depend on the sampled magnetic configurations, as the magnitude of renormalization is strongly influenced by the choice of dataset.
In fact, as discussed later in Sec.~\ref{3d}, this dependence is one of the key points of this study.

 The validity of the renormalization should be carefully justified by assessing whether the Heisenberg model sufficiently reproduces the DFT energies.
If it does not, higher-order cluster interactions should be explicitly incorporated into the spin model.
To address this point, in Sec.~S8 of the Supplementary Information~\cite{SI}, we compare the energies calculated using DFT and the Heisenberg model for the systems examined in this study.

\section{Computational conditions}
\label{conditions}
\subsection{DFT calculations in the (SC)$^2$ method}
In the (SC)$^2$ method, we employed constrained noncollinear spin DFT within the projector augmented-wave method~\cite{Blochl1994-fq} as implemented in the VASP package~\cite{Kresse1996-oa, Kresse1999-sm}.
The constrained local moment approach was employed to fix the direction of the atomic magnetic moments, while the magnitudes of the magnetic moments were allowed to relax.
As the exchange-correlation functional, we used two generalized gradient approximations: PBEsol~\cite{Perdew2008-yc} for SrMnO$_3$, and PBE~\cite{Perdew1996-sf} for Nd-based permanent-magnet compounds and elemental 3$d$ transition metals.
These choices were made to facilitate comparisons with previous studies~\cite{Zhu2020-ik, Harashima2021-iz}.
The energy convergence criterion was set to $10^{-7}$ eV/atom.
The spin-orbit coupling was not incorporated throughout this study.
Other system-dependent conditions are described separately below.

We discuss the validity of relaxing the magnitudes of the magnetic moments. 
While classical spin models, including Eqs.~(\ref{heisenberg}) and (\ref{eq:sce}), focus solely on the degrees of freedom associated with the spin orientations, 
this does not necessarily imply that the magnitudes of the magnetic moments remain fixed in DFT calculations.
Rather, the contributions of longitudinal variations in magnetic moments are renormalized into the $J_{ij}$ parameters through the regression procedure in Eq.~(\ref{least-squares}).
Therefore, relaxing the magnitudes of the magnetic moments is appropriate as long as the contribution of longitudinal spin fluctuations remains insignificant.

\subsubsection{SrMnO$_3$ with the type--G antiferromagnetic state}
The plane-wave cutoff was set to 680 eV, and a $k$-grid spacing of approximately 0.17 \AA$^{-1}$ along each reciprocal direction was used.
To incorporate on-site Coulomb interactions, we applied the DFT+$U$ method~\cite{Dudarev1998-ic} using $U_{\rm eff} =$ 3.0 eV~\cite{Zhu2020-ik}.
We used two lattice constants following Ref.~\cite{Zhu2020-ik}, specifically $a/a_{\rm eq} =$ 1.0 and 1.05, where the equilibrium lattice constant $a_{\rm eq} = $ 3.79 \AA\ was adopted.
A $2\times2\times2$ cubic cell (containing 40 atoms) was used.
The radii of the atomic spheres, to which the constrained magnetic field was applied (VASP parameter RWIGS), were set to 2.0 \AA\ for Sr, 1.2 \AA\ for Mn, and 0.7 \AA\ for O.

\subsubsection{Nd$_2$Fe$_{14}$B and Nd$_2$Co$_{14}$B}
The plane-wave cutoff was set to 350 eV, and a $k$-grid spacing of approximately 0.18 \AA$^{-1}$ along each reciprocal direction was used.
Lattice constants were set to experimental values~\cite{Le-Roux1985-wq, Herbst1985-yh}, whereas internal atomic coordinates were optimized.
The radii of the atomic spheres for the constrained magnetic field were set as 1.5 \AA\ for Nd, 1.2 \AA\ for Fe and Co, and 0.8 \AA\ for B.
We used a $1\times1\times1$ unit cell (containing 68 atoms).
Although this relatively small unit cell introduces a large finite-size effect in the evaluation of $J_{ij}$ itself, it does not pose major issues for the evaluation of $T_{\rm C}$ using the mean-field approximation~\cite{Matsubara1977-jf} employed later: the size effect causes interactions with distant equivalent atomic pairs to be incorporated into the $J_{ij}$ of nearby pairs, whereas the mean-field approximation eventually sums all $J_{ij}$ between equivalent pairs.

\subsubsection{Elemental 3$d$ transition metals with bcc and fcc structures}
The plane-wave cutoff was set to 350 eV, and a $k$-grid spacing of approximately 0.18 \AA$^{-1}$ along each reciprocal direction was used.
We used $4\times4\times4$ cubic supercells for bcc and $3\times3\times3$ cubic supercells for fcc.
The radii of the atomic spheres were set to half the nearest neighbor distance in each system.
The dependence of $J_{ij}$ on the radii of the atomic spheres and the convergence with respect to the supercell size were summarized in Secs.~S4 and S5 of the Supplementary Information~\cite{SI}.
The lattice constants, reference magnetic configurations, and magnetic moments within an atomic site are provided in Table~\ref{params}.
Note the following two points: for bcc Cr, the reference magnetic state was assumed to be the type--I antiferromagnetic state in our calculations, although a spin-density wave is observed in experiments; 
for fcc Mn, we adopted a relatively large lattice constant (3.86 \AA) to obtain a ferromagnetic state, because this system exhibits a nonmagnetic state in a wide range of volumes around the equilibrium point.

\begin{table}
\caption{Lattice constants ($a$), reference magnetic configurations, and the magnitude of magnetic moment $|\mathfrak{m}|$ per atomic site of elemental 3$d$ transition metals. The notations AFM and FM denote antiferromagnetic and ferromagnetic, respectively.
Lattice constants were obtained using structure optimization, except for fcc Mn (see the main text).}
\begin{tabular}{lrcc}
 	\hline \hline
	System & $a$ (\AA) & Reference magnetic configuration & $|\mathfrak{m}|$  ($\mu_{\rm B}$)\\
	\hline
	bcc \\
	
	\quad Cr & 2.85 & Type--I AFM & 1.04 \\
	\quad Mn & 2.79 & FM & 0.89 \\
	\quad Fe & 2.83 & FM & 2.21 \\
	\quad Co & 2.80 & FM & 1.81 \\
	\quad Ni & 2.80 & FM & 0.59 \\
	\hline 
	fcc \\
	\quad Mn & 3.86 & FM & 2.98 \\
	\quad Fe & 3.64 & FM & 2.60 \\
	\quad Co & 3.51 & FM & 1.65 \\
	\quad Ni & 3.52 & FM & 0.62 \\
	\hline 
\end{tabular}
\label{params}
\end{table}

\subsection{MFT-based method}
The family of MFT-based methods has been implemented using various basis sets and DFT packages.
To minimize the dependence on basis sets, we adopted tight-binding models with Wannier functions constructed from the same VASP package~\cite{Kresse1996-oa, Kresse1999-sm} in the case of the (SC)$^2$ method.
The Wannier functions were generated using the Wannier90 package~\cite{Pizzi2020-gj}, and the TB2J package~\cite{He2021-dw} was subsequently employed to evaluate $J_{ij}$.
Section S6 of the Supplementary Information~\cite{SI} presents a comparison of band dispersions obtained from the tight-binding model using Wannier functions and those calculated from DFT.

Unfortunately, in the case of Nd-based magnet compounds, the evaluations of $J_{ij}$ via constructing Wannier functions are very computationally demanding.
We therefore employed the KKR Green's function method within the muffin-tin approximation implemented in the AkaiKKR package~\cite{Akai1989-yk}.
The 4$f$ electrons in the Nd atoms were treated as core electrons (open-core approximation).
The maximum angular momentum of partial waves for all elements was set to 2.
As reference magnetic states, we used both the ferromagnetic state and the local moment disorder (LMD) state, in which equal fractions of up- and down-spin-polarized atoms are treated within the coherent potential approximation, yielding zero net magnetization while retaining finite local moments~\cite{Akai1993-hs}.
The same structural parameters used in the (SC)$^2$ method were applied in all cases.
To verify that the converged electronic states obtained from both packages (VASP and AkaiKKR) are sufficiently consistent, we compared the magnetic moments of all sites, as summarized in Sec.~S2 of the Supplementary Information~\cite{SI}.

\subsection{Classical Monte Carlo simulations}
Classical Monte Carlo simulations, implemented in the EspinS package~\cite{Rezaei2022-rv}, were carried out to numerically evaluate the magnetic phase transition temperatures of elemental 3$d$ transition metals.
Interaction shells up to the 10th nearest neighbor for bcc structures and the 7th nearest neighbor for fcc structures were considered for $J_{ij}$.
The results of the (SC)$^2$ method were calculated using $J_{ij}$ in a $4\times4\times4$ supercell for bcc and a $3\times3\times3$ supercell for fcc.
For both structures, $16\times16\times16$ sites were adopted, with 300,000 steps for equilibration and 2,000,000 steps for sampling.
The magnetic phase transition temperatures were defined as the temperature at which the specific heat exhibits a peak.

\section{Results}
\label{results}

\subsection{SrMnO$_3$ with the type--G antiferromagnetic state}
\label{srmno3}
Perovskite SrMnO$_3$ exhibits multiferroicity~\cite{Edstrom2018-lx} and has therefore attracted interest for new information technologies.
In addition, the strong dependence of $J_{ij}$ between Mn atoms on strain, which induces magnetic phase transitions, was experimentally examined~\cite{Maurel2015-ku}.
Thus, the quantitative evaluation of $J_{ij}$ is crucial for reliable materials and device design with sensitive magnetic properties.
A recent study, however, demonstrated that the $J_{ij}$ of SrMnO$_3$ obtained within MFT differs qualitatively from the results of self-consistent calculations~\cite{Zhu2020-ik};
under isotropic volume expansion, the MFT-based method yields a negative first-nearest-neighbor exchange coupling parameter $J_{01}$ in the type--G antiferromagnetic state (antiferromagnetic-favoring), whereas self-consistent calculations give a positive value (ferromagnetic-favoring).
Regarding the sign of $J_{01}$, geometric considerations of the superexchange mechanism (the Kanamori-Goodenough-Anderson rule) suggest that the linear Mn--O--Mn arrangement should favor antiferromagnetic coupling, as noted in Refs.~\cite{Sondena2006-gh, Edstrom2018-lx, Zhu2020-ik}.
However, the sign reversal of $J_{01}$ under volume expansion does not appear to be readily explained within this framework, which makes the phenomenon particularly intriguing.
Here, we reexamine the results of Ref.~\cite{Zhu2020-ik} and discuss the origin of the difference in $J_{01}$ in terms of the electron--spin coupling.

\begin{figure}
\centering
\includegraphics[width=70mm]{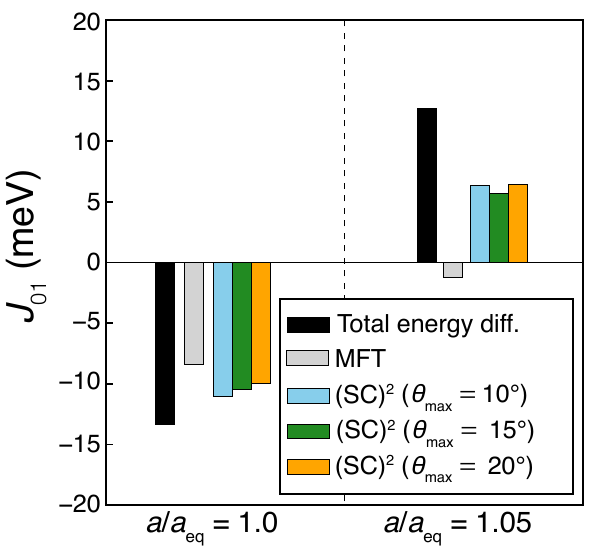}
\caption{
Exchange coupling parameters at the first-nearest neighbor Mn--Mn pair $J_{01}$ in cubic SrMnO$_3$ with the type--G antiferromagnetic state.
The black and gray bars represent the results obtained from the total-energy method in Eq.~(\ref{ediff}) and the MFT-based method, respectively.
The other colored bars show the results from the (SC)$^2$ method with three different upper bounds for the polar angle $\theta$: 10$^\circ$ (blue), 15$^\circ$ (green), and 20$^\circ$ (orange).
The lattice constants were set to $a/a_{\rm eq} =$ 1.0 and 1.05, where $a_{\rm eq} =$ 3.79 \AA.
}
\label{srmno3_j01}
\end{figure}

\begin{figure}
\centering
\includegraphics[width=80mm]{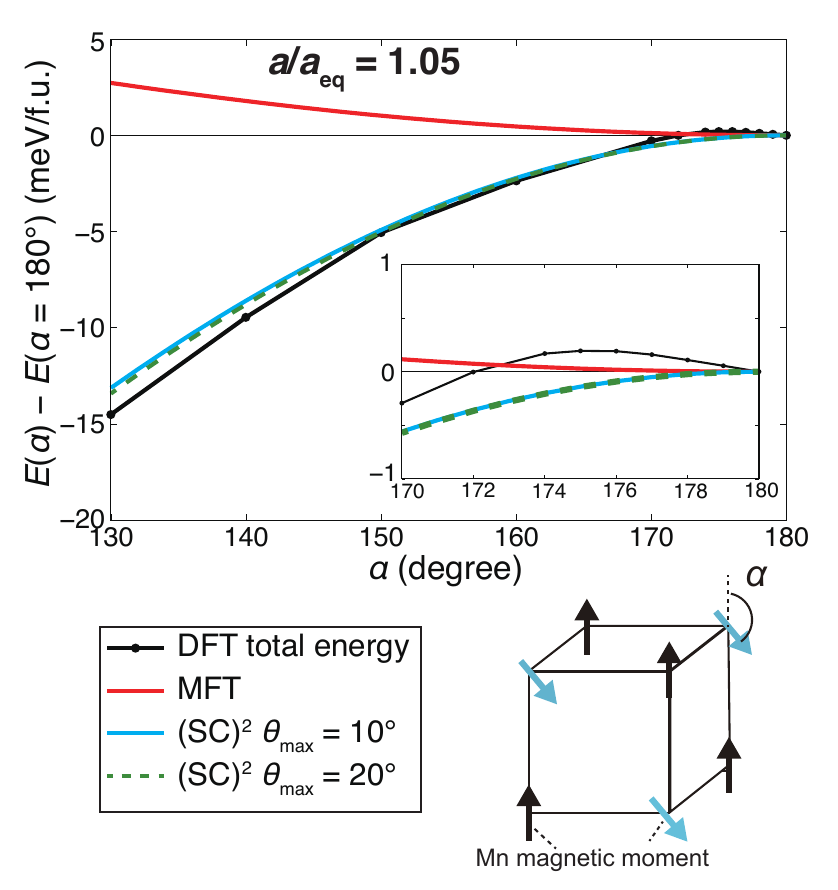}
\caption{
Total-energy variation (per formula unit) for the lattice constant $a/a_{\rm eq} = 1.05$ as a function of the rotation angle $\alpha$ of Mn magnetic moments, illustrated in the lower-right.
The type--G antiferromagnetic state corresponds to $\alpha = 180^\circ$.
The black line represents DFT total energy.
The blue, dashed green, and red lines depict the energy variations using Eq.~(\ref{energy_variation}) with $J_{ij}$ obtained from the (SC)$^2$ method for $\theta_{\rm max} = 10^\circ$, $\theta_{\rm max} = 20^\circ$, and the MFT-based method, respectively.
The inset shows a magnified view of the energy variations in the range between $170^\circ$ and $180^\circ$.
}
\label{srmno3_energy}
\end{figure}

\begin{figure*}
\centering
\includegraphics[width=160mm]{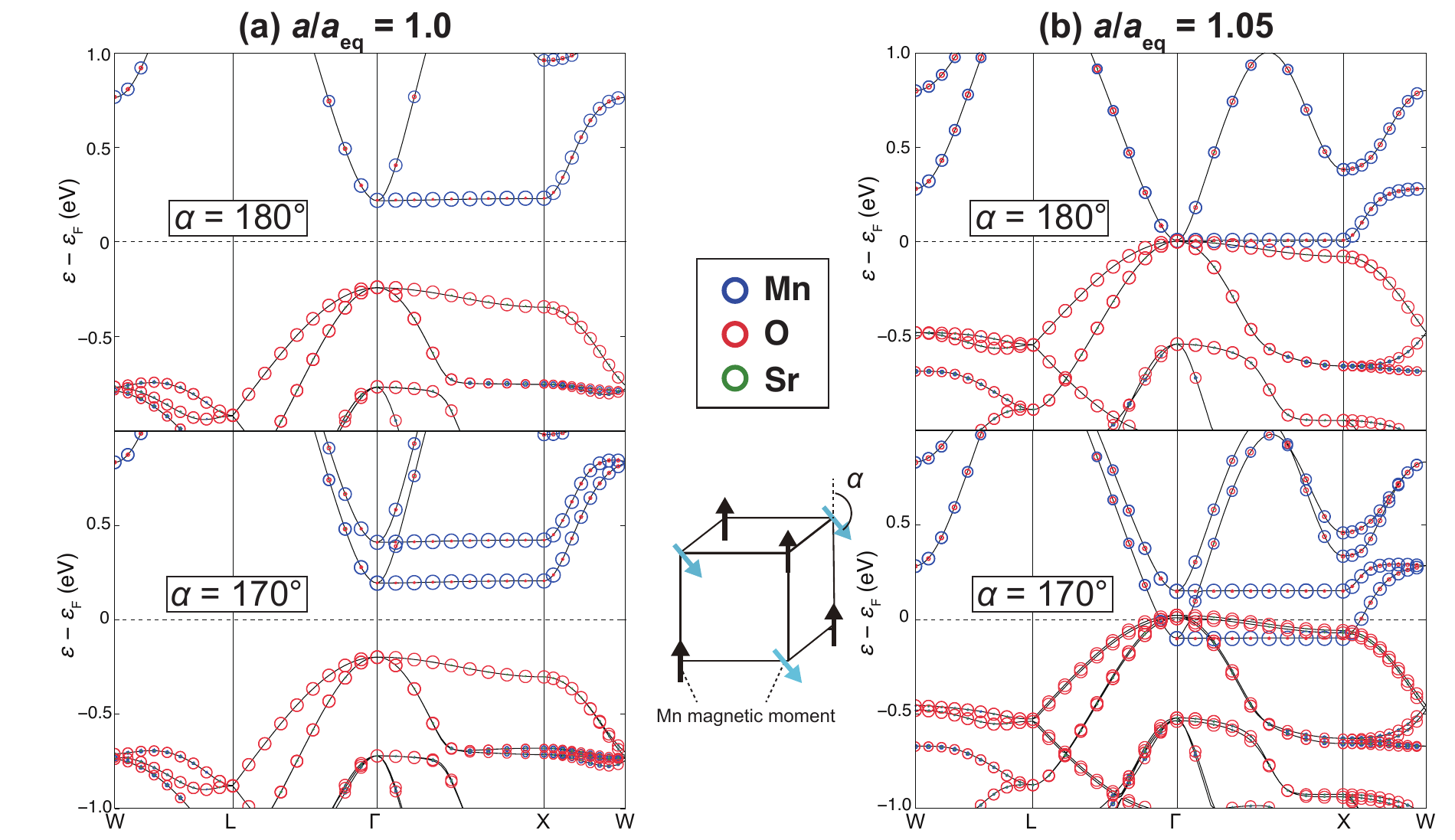}
\caption{
Element-projected band dispersions of SrMnO$_3$ with and without rotation of Mn magnetic moments, as illustrated in the inset.
Panel (a) corresponds to the equilibrium lattice constant $a/a_{\rm eq} = 1.0$, whereas panel (b) corresponds to $a/a_{\rm eq} = 1.05$.
In each panel, the upper and lower plots correspond to $\alpha=180^\circ$ and $170^\circ$, respectively.
The blue, red, and green circles denote the contributions from Mn, O, and Sr, respectively.
The band dispersions and the element contributions were plotted using VASPKIT~\cite{Wang2021-vv}.
}
\label{srmno3_band}
\end{figure*}

Figure~\ref{srmno3_j01} shows $J_{01}$ between Mn atoms in cubic SrMnO$_3$ for different lattice constants $a/a_{\rm eq} =$ 1.0 and 1.05 (the $J_{ij}$ values for more distant atomic pairs are summarized in Sec.~S1 of the Supplementary Information~\cite{SI}).
In addition to the results from the MFT-based and the (SC)$^2$ methods, we also present results from a total-energy method, where $J_{01}$ is approximated from the energy difference between the type--G antiferromagnetic and ferromagnetic states as follows:
\begin{align}
J_{01} = \frac{E^{\rm AFM}_{\rm total} - E^{\rm FM}_{\rm total}}{N},
\label{ediff}
\end{align}
where $N$ ($= 6$ in this case) is the number of equivalent Mn--Mn bonds at the first-nearest neighbor, $E^{\rm AFM}_{\rm total}$ and $E^{\rm FM}_{\rm total}$ are the total energies of the type--G antiferromagnetic and ferromagnetic states per formula unit, respectively.
This total-energy method makes it easy to see the relative energy difference between the two magnetic states ($J_{ij}$ at the second-nearest neighbor and beyond are much smaller than those at the first-nearest neighbor).
As reported in Ref.~\cite{Zhu2020-ik}, the MFT-based method yields a negative $J_{01}$ (favoring antiferromagnetic coupling) for both lattice constants.
However, the total-energy method indicates that $J_{01}$ for $a/a_{\rm eq} =$ 1.05 is positive.
On the other hand, the (SC)$^2$ method correctly reproduced the change in the sign of $J_{01}$ with volume expansion.

To investigate the differences between the MFT-based method and the (SC)$^2$ method, we calculated the total-energy variation as a function of the rotation angle $\alpha$ of Mn magnetic moments, as shown in Fig.~\ref{srmno3_energy}, following Ref.~\cite{Zhu2020-ik}.
The rotation angle is continuously varied from the type--G antiferromagnetic state to the ferromagnetic state.
The Heisenberg models derived from the MFT-based method and the (SC)$^2$ method also yield energy variations per formula unit as a function of $\alpha$~\cite{Zhu2020-ik}:
\begin{align}
\begin{split}
E(\alpha) &= -\frac{1}{N_{\rm Mn}}\sum_{i\neq j}J_{ij} \hat{\bm e}_i \cdot \hat{\bm e}_j\\
&= -(6J_{01}+8J_{03})\cos\alpha,
\label{energy_variation}
\end{split}
\end{align}
where $N_{\rm Mn}$ ($= 2$ in this case) is the number of Mn atoms, and $J_{03}$ is the third-nearest-neighbor exchange coupling parameter (although its magnitude is negligible in both methods, as shown in Ref.~\cite{Zhu2020-ik} and Sec.~S1 of the Supplementary Information~\cite{SI}).
The results of the (SC)$^2$ method correctly reproduce the total-energy variation in the range from $\alpha=$ 180$^\circ$ (the type--G antiferromagnetic state) down to $\alpha=$ 140$^\circ$. In contrast, the MFT-based method seems to fail to capture the energy variation in this range, consistent with Ref.~\cite{Zhu2020-ik}.
However, as shown in the inset of Fig.~\ref{srmno3_energy}, the MFT-based method qualitatively captures the stability of the magnetic state correctly in the immediate vicinity of 180$^\circ$. In other words, the sign of the curvature of the energy landscape at 180$^\circ$ is consistent.
Narrowing the sampling window of the (SC)$^2$ method to $\theta_{\rm max}=3^\circ$ does not recover the local minimum at $180^\circ$ (Sec.~S1 of the Supplementary Information~\cite{SI}).
The fitted $J_{ij}$ for $\theta_{\rm max}=3^\circ$ differ markedly from those for $10^\circ$ and $20^\circ$ and do not describe the local curvature of the energy landscape at this gapless state.

\begin{figure}[h]
\centering
\includegraphics[width=\linewidth]{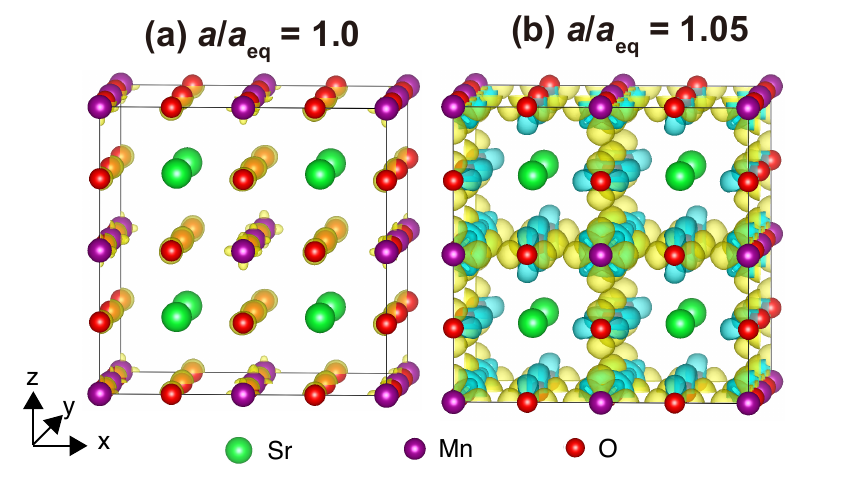}
\caption{
Charge density differences $\Delta n(\bm{r})$ in SrMnO$_3$ as defined in Eq.~(\ref{diffden}) between rotation angles $\alpha = 170^\circ$ and $\alpha = 180^\circ$, drawn using VESTA software~\cite{Momma2011-yj}.
Panel (a) corresponds to the equilibrium lattice constant $a/a_{\rm eq} = 1.0$, whereas panel (b) corresponds to $a/a_{\rm eq} = 1.05$.
The isosurface levels represented in yellow and blue are set to $+2\times10^{-5}$ and $-2\times10^{-5}$ bohr$^{-3}$, respectively.
}
\label{srmno3_chg}
\end{figure}

To examine the origin of the local minimum at $\alpha = 180^\circ$, we focus on the underlying electronic structure.
Figures~\ref{srmno3_band}(a) and (b) show the element-projected band dispersions of SrMnO$_3$ at $a/a_{\rm eq}=$ 1.0 and 1.05, respectively.
For each lattice constant, we compare the band structure for $\alpha=180^\circ$ (the type--G antiferromagnetic state) with that for a slightly rotated configuration, $\alpha=170^\circ$.
For $a/a_{\rm eq} =$ 1.0, the band gap remains open, regardless of the magnetic moment rotation (Fig.~\ref{srmno3_band}(a)).
In contrast, for $a/a_{\rm eq} =$ 1.05, the band gap closes, and flat bands consisting of Mn and O states appear near the Fermi level along the $\Gamma$--${\rm X}$ line (upper panel of Fig.~\ref{srmno3_band}(b)).
This gap closure upon volume expansion can be understood as a consequence of weakened Mn--O hybridization~\cite{Sondena2006-gh}, which reduces the bonding-antibonding splitting and brings the valence- and conduction-band edges into contact. 
Upon rotating magnetic moments, the degenerate Mn bands split into occupied and unoccupied states, while the occupied O bands near the $\Gamma$ point shift into the unoccupied region (lower panel of Fig.~\ref{srmno3_band}(b)).
Consequently, the changes in electron density associated with the rotation of Mn magnetic moments differ significantly before and after the volumetric expansion.
Figure~\ref{srmno3_chg} shows the charge density differences $\Delta n(\bm{r})$ resulting from the rotation of Mn magnetic moments for $a/a_{\rm eq} =$ 1.0 and 1.05, defined as
\begin{align}
\Delta n(\bm{r}) = n^{\alpha = 170^{\circ}}(\bm{r}) - n^{\alpha = 180^{\circ}}(\bm{r}),
\label{diffden}
\end{align}
where $n^{\alpha}(\bm{r})$ represents the electron density at position $\bm{r}$ when magnetic moments are rotated by an angle $\alpha$.
In the case of $a/a_{\rm eq} =$ 1.05 (Fig.~\ref{srmno3_chg}(b)), we find a substantial charge-density difference at the Mn and O sites compared with the case of $a/a_{\rm eq} =$ 1.0 (Fig.~\ref{srmno3_chg}(a)), which we attribute to the change in orbital occupations described above.
These spin-rotation-induced changes in the charge density reflect a strong electron--spin coupling.
Accordingly, the associated energetic effect of electron--spin coupling becomes significant and is captured only when the electronic structure is treated fully self-consistently.

Although the MFT-based method may appear to give an inconsistent result for SrMnO$_3$ (the type--G antiferromagnetic state under volume expansion), it is formally correct in the strict limit of infinitesimal spin rotations, reproducing the sign of the local energy curvature around the AFM state.
Thus, the discrepancy reported in Ref.~\cite{Zhu2020-ik} between self-consistent finite-rotation calculations and MFT primarily reflects the use of finite rotation angles, rather than a failure of MFT in the infinitesimal-rotation limit.
At finite angles, changes in band occupancy arise and contribute to electron--spin coupling; because conventional MFT neglects the accompanying self-consistent variations in the charge and magnetization densities, these effects fall outside its scope and delineate its practical domain of applicability.
In such situations, nonperturbative approaches, including the (SC)$^2$ method, are more appropriate for constructing an effective spin model over a finite region of configuration space.
The importance of electron--spin coupling might seem limited to cases near a metal--insulator transition; however, as shown below for representative metallic systems, electron--spin coupling is likewise important for the quantitative evaluation of the effective $J_{ij}$.

\subsection{Nd$_2$Fe$_{14}$B and Nd$_2$Co$_{14}$B}
\label{2-14-1}
Evaluation of $J_{ij}$ plays an important role in the materials design of magnet compounds, because the stronger exchange couplings suppress demagnetization due to temperature, i.e., larger $J_{ij}$ values lead to higher $T_{\rm C}$.
Nevertheless, previous methods based on the MFT for evaluating $T_{\rm C}$ through $J_{ij}$ of magnet compounds have been unsatisfactory in terms of quantitative accuracy;
the MFT-based method failed to reproduce the increase in $T_{\rm C}$ when Fe was substituted with Co in Nd$_2$Fe$_{14}$B~\cite{Harashima2021-iz, Miyake2021-rl}.
Magnetic interactions are important not only for understanding existing permanent magnets~\cite{Miyake2021-rl, Gohda2021-ek} but also for exploring new magnetic compounds~\cite{Seo2024-nd}.
This problem, therefore, must be resolved for the reliable materials design of magnet compounds.

\begin{figure}
\centering
\includegraphics[width=\linewidth]{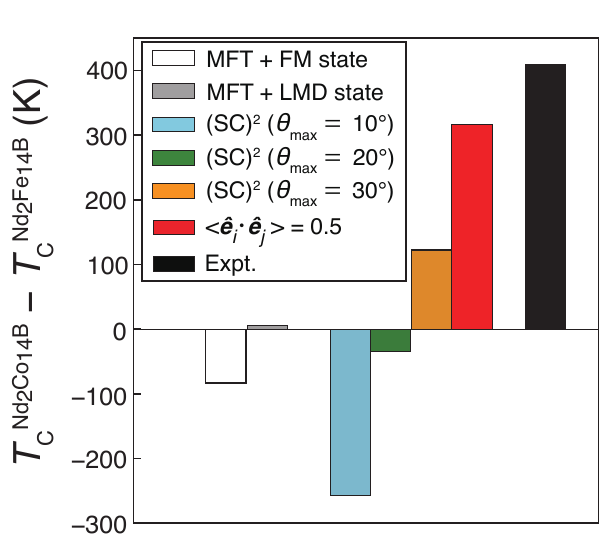}
\caption{
Differences in $T_{\rm C}$ of Nd$_2$Co$_{14}$B and Nd$_2$Fe$_{14}$B.
The white and gray bars show MFT-based results obtained from FM and LMD reference states, respectively.
The black bar denotes experimental values.
The blue, green, and orange bars indicate (SC)$^2$ results with the maximum rotation angle set to $\theta_{\rm max} = 10^\circ$, $20^\circ$, and $30^\circ$, respectively.
The red bar indicates the value obtained by the (SC)$^2$ method using the spin-direction distribution in Eq.~(\ref{spin_dist}) with $\tau=0.6261$, chosen to reproduce the nearest-neighbor spin--spin correlation at the Curie temperature of bcc Fe reported in previous theoretical studies~\cite{Melnikov2019-es, Walsh2022-ci}, $\langle \hat{\bm e}_i \cdot \hat{\bm e}_j \rangle = 0.5$.
All theoretical values are obtained using the mean-field approximation.
Experimental values are taken from Refs.~\cite{Sagawa1984-rp, Fuerst1988-qm}.
}
\label{tc_2-14-1}
\end{figure}

Figure~\ref{tc_2-14-1} shows the differences between $T_{\rm C}$ values of Nd$_2$Fe$_{14}$B and Nd$_2$Co$_{14}$B.
In the MFT-based method, we used FM and LMD states as reference states and obtained results in line with previous studies~\cite{Harashima2021-iz, Miyake2021-rl}.
However, as in those works, the MFT-based calculations do not reproduce the increase in $T_{\rm C}$ upon substituting Fe with Co, irrespective of whether the FM or LMD reference is used.
Although a similar disagreement with experiment is also seen in the (SC)$^2$ results at $\theta_{\rm max}=10^\circ$, the correct trend emerges as $\theta_{\rm max}$ increases;
$T_{\rm C}$ of Nd$_2$Co$_{14}$B becomes larger than that of Nd$_2$Fe$_{14}$B when $\theta_{\rm max}=30^\circ$.
Finally, when we use the spin-direction distribution in Eq.~(\ref{spin_dist}) tuned to yield a nearest-neighbor correlation of $\langle \hat{\bm e}_i \cdot \hat{\bm e}_j \rangle = 0.5$, the calculated $T_{\rm C}$ difference comes much closer to the experimental value.
This indicates that the reduction of $T_{\rm C}$ with increasing magnetic disorder is stronger for Nd$_2$Fe$_{14}$B than for Nd$_2$Co$_{14}$B (absolute values of $T_{\rm C}$ are summarized in Sec.~S3 of the Supplementary Information~\cite{SI}).
This disorder dependence of $T_{\rm C}$ (and of the extracted $J_{ij}$) suggests that contributions from higher-order cluster interactions beyond the Heisenberg model are effectively folded into the apparent $J_{ij}$ parameters as discussed in Sec.~\ref{sce}.
In Sec.~\ref{3d}, we show for the elemental 3$d$ metals that such a $\theta_{\rm max}$ dependence is not primarily captured within a band-energy-only, frozen-potential picture as assumed in conventional MFT-based schemes.
Rather, it is predominantly associated with the self-consistent changes in the electronic structure induced by spin rotations, which are inherently captured only in nonperturbative treatments.

In finite-temperature regimes with moderate spin fluctuations, a nonperturbative treatment that explicitly includes electron--spin coupling can provide a complementary description.
We acknowledge that variants of MFT-based approaches (e.g., including constraining fields~\cite{Bruno2003-yi, Katsnelson2004-pw, Solovyev2021-bz}, embedding electronic correlations~\cite{Katsnelson2000-nm, Pourovskii2016-oc, Katanin2023-cn, Katanin2023-dl}, or considering longitudinal spin fluctuations~\cite{Ruban2007-ae}) may reproduce the composition dependence of the transition temperatures in Nd$_2$Fe$_{14}$B and Nd$_2$Co$_{14}$B.
Even so, the finite-angle (and thus temperature) dependence revealed by the (SC)$^2$ method highlights that electron--spin coupling can play a significant quantitative role, and that incorporating this feedback systematically within perturbative mappings can be challenging.
From this perspective, the (SC)$^2$ method offers an alternative route: it targets partially disordered configurations and treats the associated electronic feedback explicitly, rather than restricting the mapping to an infinitesimal-rotation framework.

Other temperature effects may also contribute to the remaining disagreement in $T_{\rm C}$ between theory and experiments.
For example, Nd$_2$Fe$_{14}$B is reported to exhibit strong magnetism-dependent phonons~\cite{Tsuna2023-fa}, which are shifts of phonon frequencies due to magnetic disordering.
As we previously pointed out in Ref.~\cite{Tanaka2020-vf}, the magnetism-dependent phonons~\cite{Tanaka2020-qe} affect the determination of equilibrium magnetic states, thereby leading to changes in the theoretical $T_{\rm C}$.
In addition, the direct impact of lattice vibrations on $J_{ij}$ would affect $T_{\rm C}$~\cite{Heine2021-qe}.
Although further investigation incorporating additional temperature effects is necessary, our results clearly demonstrate the importance of nonperturbative methods in capturing the concentration dependence of $T_{\rm C}$ in the Nd$_2$(Fe,Co)$_{14}$B system.

\subsection{Elemental 3$d$ transition metals with bcc and fcc structures}
\label{3d}

\begin{figure*}
\centering
\includegraphics[width=160mm]{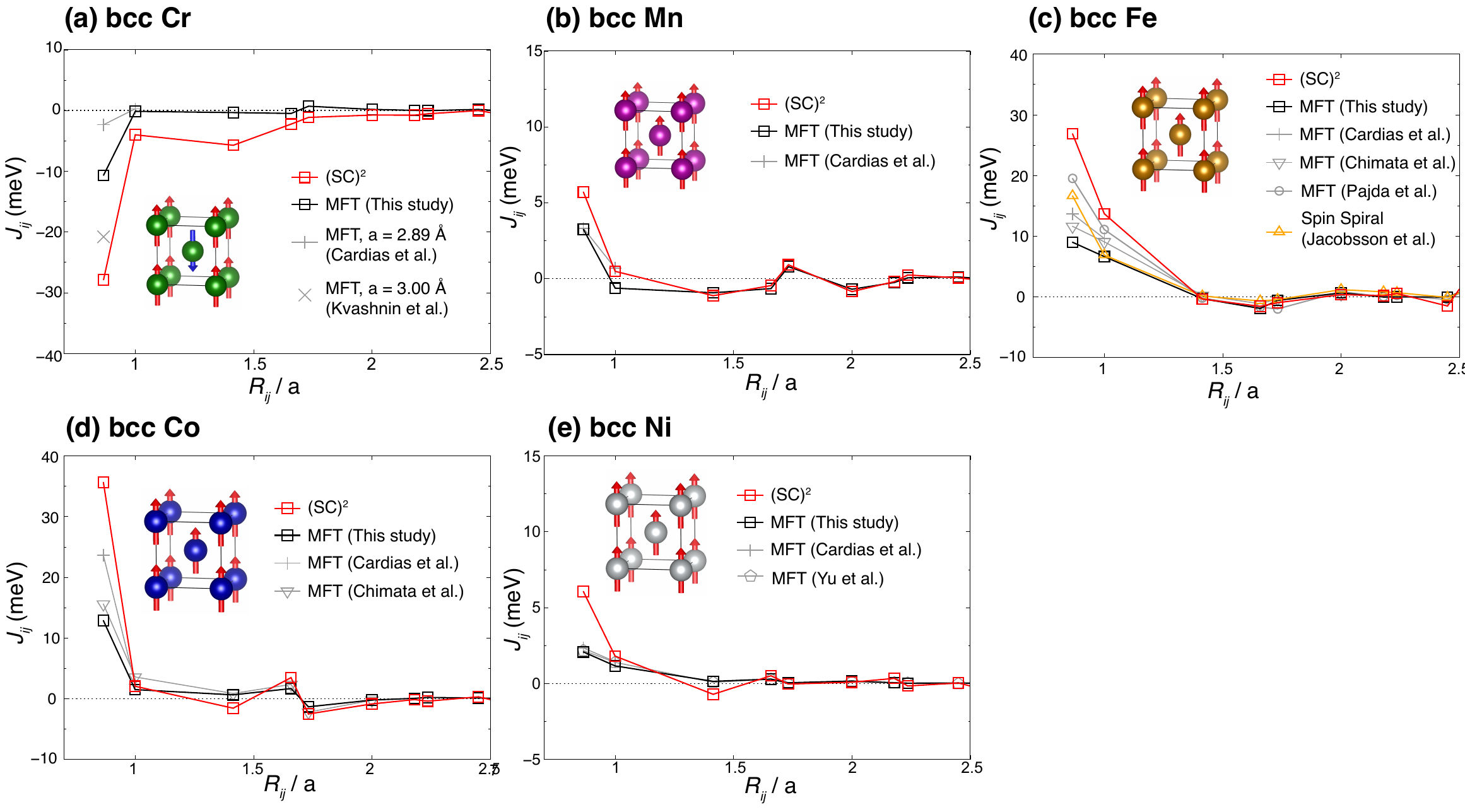}
\caption{
Exchange coupling parameters $J_{ij}$ in (a) bcc Cr, (b) bcc Mn, (c) bcc Fe, (d) bcc Co, and (e) bcc Ni as a function of the interatomic distance $R_{ij}$ (in units of the lattice constant $a$).
The results of the (SC)$^2$ method were obtained using $\theta_{\rm max} = 10^\circ$.
In the (SC)$^2$ method, a $4\times4\times4$ body-centered-cubic supercell containing 128 atoms was used.
For comparison, literature results from MFT-based methods referenced to the ground state~\cite{Cardias2017-kh, Kvashnin2016-bl, Chimata2017-du, Pajda2001-mr, Yu2008-sv} and from self-consistent spin-spiral calculations near the ferromagnetic state~\cite{Jacobsson2022-nu} are also shown. 
The reference magnetic state of each system is illustrated using VESTA software~\cite{Momma2011-yj}.
}
\label{jijbcc}
\end{figure*}

\begin{figure*}
\centering
\includegraphics[width=125mm]{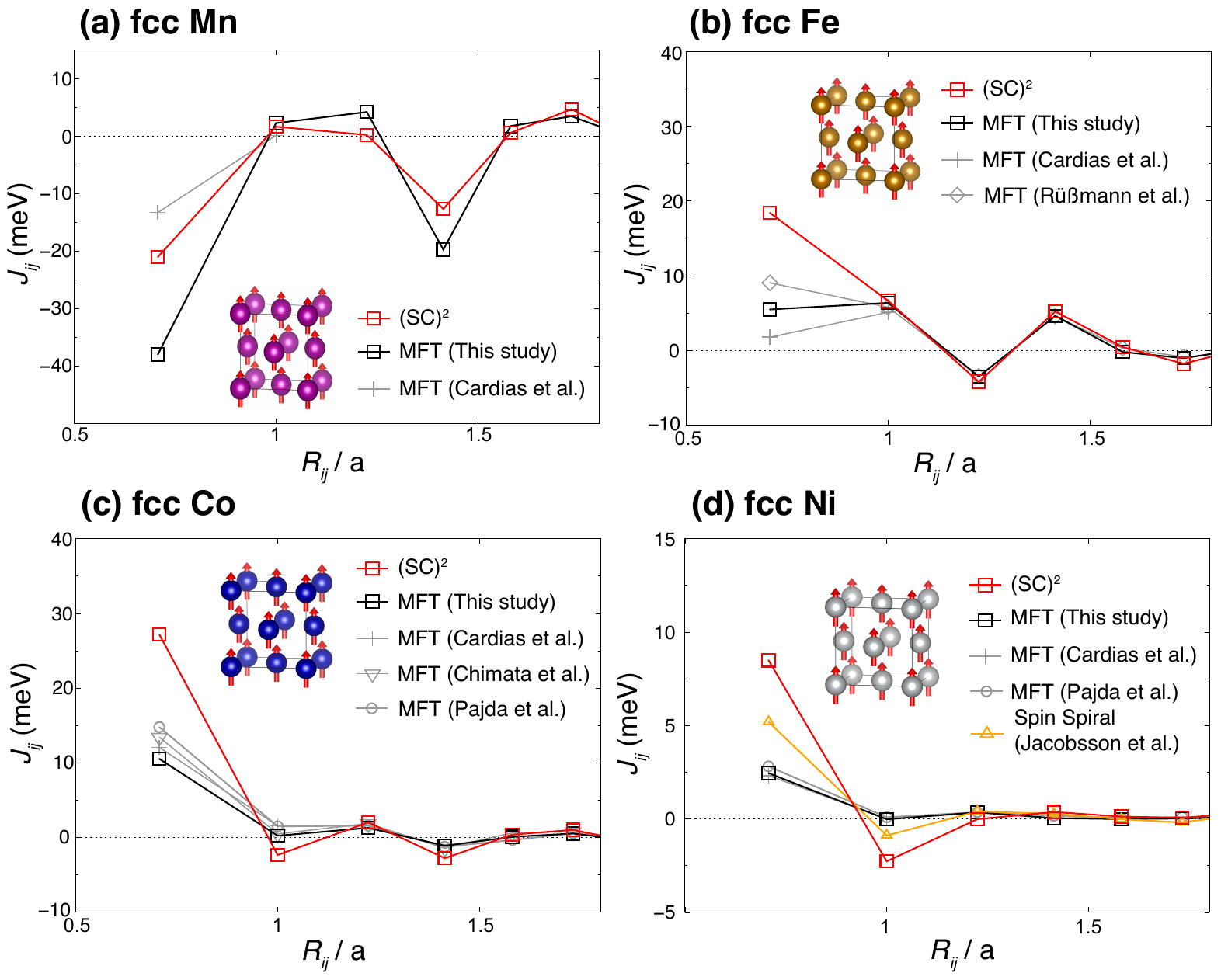}
\caption{
Exchange coupling parameters $J_{ij}$ in (a) fcc Mn, (b) fcc Fe, (c) fcc Co, and (d) fcc Ni as a function of the interatomic distance $R_{ij}$ (in units of the lattice constant $a$).
The results of the (SC)$^2$ method were obtained using $\theta_{\rm max} = 10^\circ$.
In the (SC)$^2$ method, a $3\times3\times3$ face-centered-cubic supercell containing 108 atoms was used.
For comparison, literature results from MFT-based methods referenced to the ground state~\cite{Cardias2017-kh, Russmann2022-wj, Chimata2017-du, Pajda2001-mr, Yu2008-sv} and from self-consistent spin-spiral calculations near the ferromagnetic state~\cite{Jacobsson2022-nu} are also shown.
The reference magnetic state of each system is illustrated using VESTA software~\cite{Momma2011-yj}.
}
\label{jijfcc}
\end{figure*}

Figures~\ref{jijbcc}(a)--(e) and \ref{jijfcc}(a)--(d) show the $J_{ij}$ values for 3$d$ transition metals with bcc and fcc structures, respectively, together with literature values for comparison~\cite{Cardias2017-kh, Kvashnin2016-bl, Russmann2022-wj, Chimata2017-du, Pajda2001-mr, Yu2008-sv, Jacobsson2022-nu}.
Generally, $J_{ij}$ values beyond the second-nearest neighbor exhibit relatively good agreement between the (SC)$^2$ method and MFT-based methods.
However, for most systems (with the exception of bcc Mn) we find substantial differences in $J_{01}$ between MFT-based estimates and the (SC)$^2$ method.
As a side remark, although both our data and those of Cardias et al.~\cite{Cardias2017-kh} were obtained within an MFT-based framework, a substantial discrepancy is observed for $J_{01}$ in fcc Mn.
This is probably because, as Ref.~\cite{Jing2003-yb} reports, the magnetic ground state of fcc Mn is highly sensitive to the lattice parameter; consequently, the extracted exchange couplings $J_{ij}$ can vary substantially with computational setup.
Therefore, at present, a direct comparison between the (SC)$^2$ method and MFT-based results for fcc Mn is unreliable.
Accordingly, we do not pursue fcc Mn further and analyze the origin of the large discrepancies for the remaining systems.

\begin{figure}
\centering
\includegraphics[width=60mm]{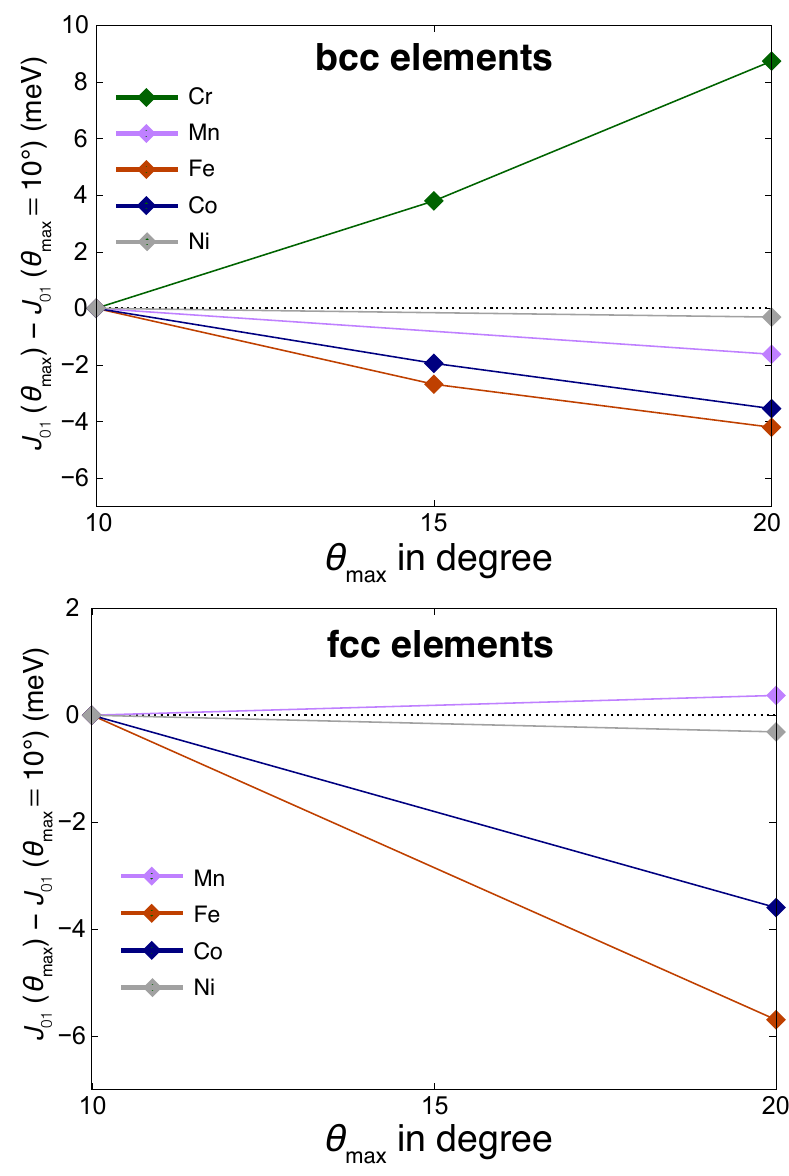}
\caption{
Dependence of $J_{01}$ on $\theta_{\rm max}$ for elemental 3$d$ transition metals.
The vertical axis represents the deviation of $J_{01}$ from its value at $\theta_{\rm max} = 10^\circ$.
}
\label{change_jij}
\end{figure}

We calculated the change in $J_{01}$ as the maximum rotation angle $\theta_{\rm max}$ was varied from 10$^\circ$ to 20$^\circ$, as shown in Fig.~\ref{change_jij}.
Systems with a large difference in $J_{ij}$ between the (SC)$^2$ method and the MFT-based method tend to also have a large angular dependence of $J_{01}$.
As discussed in Sec.~\ref{sce}, the angular dependence of $J_{01}$ (i.e., the dependence on the sampled magnetic configurations) results from the renormalization of higher-order cluster interactions beyond the Heisenberg model, which can be enhanced by electron--spin coupling through spin-dependent electronic-structure changes.
Here, we illustrate this renormalization mechanism using the biquadratic interaction ($-B_{ij}(\hat{{\bm e}}_{i} \cdot \hat{\bm e}_{j})^2$) as a representative higher-order term.
The spin Hamiltonian is
\begin{align}
{\mathcal H}= -J_{ij} \hat{{\bm e}}_{i} \cdot \hat{\bm e}_{j} - B_{ij}(\hat{{\bm e}}_{i} \cdot \hat{\bm e}_{j})^2.
\end{align}
The energy as a function of $\theta_{\rm max}$ is
\begin{align}
E = -J_{ij} \langle\hat{{\bm e}}_{i} \cdot \hat{\bm e}_{j}\rangle_{\theta_{\rm max}} - B_{ij}\langle(\hat{{\bm e}}_{i} \cdot \hat{\bm e}_{j})^2\rangle_{\theta_{\rm max}},
\end{align}
where $\langle \cdots \rangle_{\theta_{\rm max}}$ denotes an expectation value when the upper bound of $\theta$ is $\theta_{\rm max}$.
The angle-dependent $J_{ij}$, expressed as ${J_{ij}}(\theta_{\rm max})$, arises when the biquadratic term is renormalized as
\begin{align}
\begin{split}
&-J_{ij} \langle\hat{{\bm e}}_{i} \cdot \hat{\bm e}_{j}\rangle_{\theta_{\rm max}} - B_{ij}\langle(\hat{{\bm e}}_{i} \cdot \hat{\bm e}_{j})^2\rangle_{\theta_{\rm max}} \\
=& -\left( J_{ij} + B_{ij}\frac{\langle(\hat{{\bm e}}_{i} \cdot \hat{\bm e}_{j})^2\rangle_{\theta_{\rm max}}}{\langle\hat{{\bm e}}_{i} \cdot \hat{\bm e}_{j}\rangle_{\theta_{\rm max}}}\right) \langle\hat{{\bm e}}_{i} \cdot \hat{\bm e}_{j}\rangle_{\theta_{\rm max}} \\
=&-{J_{ij}}(\theta_{\rm max}) \langle\hat{{\bm e}}_{i} \cdot \hat{\bm e}_{j}\rangle_{\theta_{\rm max}},
\end{split}
\end{align}
where
\begin{align}
{J_{ij}}(\theta_{\rm max}) = J_{ij} + B_{ij}\frac{\langle(\hat{{\bm e}}_{i} \cdot \hat{\bm e}_{j})^2\rangle_{\theta_{\rm max}}}{\langle\hat{{\bm e}}_{i} \cdot \hat{\bm e}_{j}\rangle_{\theta_{\rm max}}},
\end{align}
and
\begin{align}
\frac{\langle(\hat{{\bm e}}_{i} \cdot \hat{\bm e}_{j})^2\rangle_{\theta_{\rm max}}}{\langle\hat{{\bm e}}_{i} \cdot \hat{\bm e}_{j}\rangle_{\theta_{\rm max}}}  
&= 
\frac{2}{\sin^4\theta_{\rm max}} \notag\\
&\times \left[ \left( -\frac{3}{4}\cos\theta_{\rm max} + \frac{1}{12}\cos3\theta_{\rm max} + \frac{2}{3}\right)^2 \right. \notag\\
&\ + \left. \frac{2}{9}\left(1-\cos^3\theta_{\rm max}\right)^2 \right].
\end{align}
The derivation of $\langle(\hat{{\bm e}}_{i} \cdot \hat{\bm e}_{j})^2\rangle_{\theta_{\rm max}}/\langle\hat{{\bm e}}_{i} \cdot \hat{\bm e}_{j}\rangle_{\theta_{\rm max}}$ is described in Sec.~S7 of the Supplementary Information~\cite{SI}.
The biquadratic interaction can also be evaluated within the MFT.
However, the typical values of $|B_{ij}|$ reported in such approaches ($\sim 5$ meV or $16$ meV for bcc Fe~\cite{Mankovsky2020-ej, Szilva2013-of}, $\le 0.1$ meV for hcp Co and fcc Ni~\cite{Mankovsky2020-ej}) cannot fully account for the pronounced $\theta_{\rm max}$ dependence of $J_{ij}(\theta_{\rm max})$ shown in Fig.~\ref{change_jij};
for example, if the change in $|{J_{ij}}(\theta_{\rm max})|$ when $\theta_{\rm max}$ increases from $10^\circ$ to $20^\circ$ is $1$ meV, the above estimate would require $|B_{ij}|\approx 24$ meV.
It is also unlikely that other higher-order cluster interactions obtained within perturbative expansions are sufficient to explain the observed angular dependence~\cite{Mankovsky2020-ej}.
These considerations suggest that the strong $\theta_{\rm max}$ dependence found in the (SC)$^2$ method is difficult to rationalize solely in terms of perturbative higher-order terms, and is instead consistent with an additional renormalization channel associated with electron--spin coupling.
In recent years, nonperturbative evaluations of biquadratic interactions have been actively pursued across a range of magnetic materials, and several studies report cases where $B_{ij}$ is non-negligible compared with $J_{ij}$~\cite{Zhu2016-ke, Kartsev2020-vh}.
It may also be interesting to revisit previously studied systems from the viewpoint that electron--spin coupling can provide a microscopic route to large effective higher-order terms.

To examine whether the $\theta_{\rm max}$ dependence indeed arises from electron--spin coupling rather than from the band energy at fixed density, we fitted the Heisenberg model to frozen-potential energies of the same sampled configurations, in which the charge density and the magnetization magnitude of the ferromagnetic reference state are held fixed and only the magnetization direction is rotated onto the sampled spin directions (Sec.~S12 of the Supplementary Information~\cite{SI}).
Between $\theta_{\rm max}=10^\circ$ and $20^\circ$, the frozen-potential $J_{01}$ changes by at most 2\% for bcc Fe and fcc Co, whereas the self-consistent $J_{01}$ decreases by 16\% and 13\%, respectively; the $\theta_{\rm max}$ dependence therefore originates predominantly from the self-consistent electronic response to the spin rotations rather than from the nonlinearity of the fixed-density band energy.

Prior perturbative studies have also examined the angular dependence of $J_{ij}$~\cite{Szilva2013-of, Kvashnin2016-bl, Cardias2017-kh, Chimata2017-du, Szilva2017-sw}.
In these works, the reference magnetic state is constructed by rotating a single spin by an angle $\theta$ away from the global magnetization axis, and the energy change under subsequent infinitesimal rotations is then evaluated.
The (SC)$^2$ results in Fig.~\ref{change_jij} are notable because, for several systems, they exhibit a stronger $\theta$ dependence than previously reported.
For bcc Fe, an earlier work indeed found a relatively large angular dependence; however, the difference between $\theta=10^\circ$ and $20^\circ$ is $\sim\!1$~meV~\cite{Chimata2017-du}, which is quantitatively inconsistent with Fig.~\ref{change_jij}.
Moreover, the near absence of angular dependence reported for bcc and fcc Co~\cite{Cardias2017-kh, Chimata2017-du} is in contrast to the results in Fig.~\ref{change_jij}.

We consider two plausible origins of the discrepancy described above.
First, prior studies assume a simplified reference state in which only a single spin is rotated~\cite{Szilva2013-of, Kvashnin2016-bl, Cardias2017-kh, Chimata2017-du, Szilva2017-sw}, whereas in the present work we emulate finite-temperature magnetism more realistically by letting all spins fluctuate simultaneously.
Compared with a single-spin-rotated reference state, such multi-spin configurations can induce a more pronounced electron--spin coupling, which may manifest as a stronger apparent angular dependence of the effective $J_{ij}$.
Second, although the previous studies compute a $\theta$-dependent $J_{ij}$ around a tilted reference state,
they effectively neglect the $\theta$ dependence of the reference-state energy itself.
In other words, those approaches construct a \emph{local} effective spin Hamiltonian:
$J_{ij}$ is extracted from energy variations around a chosen (spin-rotated) reference configuration, while the total energy of the reference state serves only as a fixed baseline.
On the other hand, our approach determines $J_{ij}$ by fitting the \emph{global} energy landscape within the sampled set of noncollinear configurations, thereby yielding effective interactions that renormalize the net impact of higher-order terms over that configuration range.
See also Refs.~\cite{dos-Santos-Dias2021-av, Cardias2022-qv, dos-Santos-Dias2022-ad, Cardias2023-hl} for discussions on the relationship between \emph{local} and \emph{global} spin Hamiltonians.
From this perspective, when one aims at quantitatively reproducing configuration-dependent energies in thermally fluctuating states, the (SC)$^2$ method may offer a practical alternative.

We comment on the fact that the $J_{ij}$ values obtained from the (SC)$^2$ method do not coincide with the MFT results in the limit $\theta_{\rm max}\to 0$.
A reasonable explanation is that the original MFT-based method used here is rigorously justified in the adiabatic long-wavelength (spin-wave) limit, whereas practical applications often require an effective description beyond this limit~\cite{Antropov2003-ph, Antropov2006-wx}.
Along these lines, several extensions of MFT have been proposed that aim to go beyond the long-wavelength formulation and to reduce systematic errors inherent in the original MFT-based mapping~\cite{Antropov2003-ph, Bruno2003-yi}.
In many cases, such extended formulations lead to a renormalization of the effective exchange scale, yielding larger exchange stiffness and higher transition temperatures than the original MFT~\cite{Bruno2003-yi, Katsnelson2004-pw, Turek2006-wy, Rajeev-Pavizhakumari2025-zw}.
This trend suggests that the magnitude of $J_{ij}$ values may also increase when using approaches that go beyond the original MFT framework, which could help reconcile the (SC)$^2$ and MFT-based results in the $\theta_{\rm max}\to 0$ limit.
Indeed, for fcc Ni, the first-nearest-neighbor exchange obtained beyond the long-wavelength approximation, $J_{01}=8.3$~meV~\cite{Antropov2003-ph}, is consistent with our (SC)$^2$ value in Fig.~\ref{jijfcc}(d).
We emphasize, however, that these extended MFT approaches still rely on an infinitesimal-rotation framework and therefore do not explicitly incorporate the electron--spin coupling that becomes relevant beyond the strict infinitesimal limit.
Finally, MFT results can also depend on the choice of basis functions and other implementation details~\cite{Szilva2023-ta, Solovyev2021-bz}.

\begin{figure}
\centering
\includegraphics[width=\linewidth]{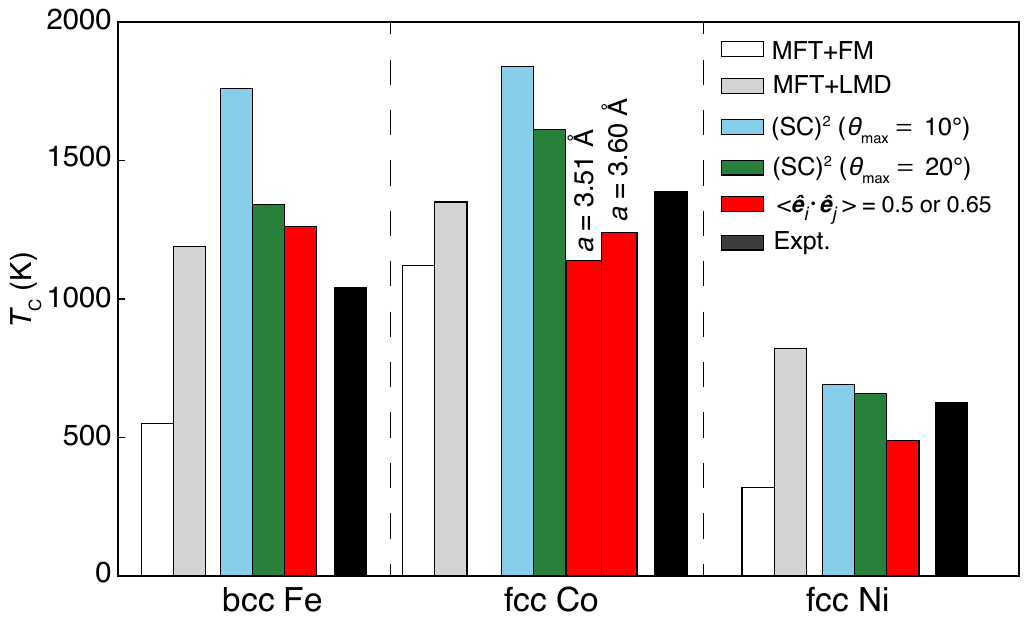}
\caption{
Comparison of $T_{\rm C}$ values for bcc Fe, fcc Co, and fcc Ni obtained by the (SC)$^2$ method and experiments~\cite{Haynes2016-rh}.
For the (SC)$^2$ results combined with the spin-direction distribution in Eq.~\eqref{spin_dist} (red bars), we used $\tau$ values chosen to reproduce the nearest-neighbor spin--spin correlation $\langle \hat{\bm e}_i \cdot \hat{\bm e}_j \rangle$ reported in previous theoretical studies~\cite{Melnikov2019-es, Walsh2022-ci}:
for bcc Fe, $\langle \hat{\bm e}_i \cdot \hat{\bm e}_j \rangle = 0.5$ ($\tau = 0.6261$), whereas for fcc Ni, $\langle \hat{\bm e}_i \cdot \hat{\bm e}_j \rangle = 0.65$ ($\tau = 0.4688$). For fcc Co, for which no corresponding reference value is available, we adopted the fcc Ni value, as both have the same crystal structure.
In addition, for fcc Co we show not only the result at the lattice constant obtained from structure optimization ($3.51$~\AA) but also the result at the experimentally reported lattice constant at $1400$~K ($3.60$~\AA)~\cite{Nishizawa1983-zh}, to approximately account for thermal expansion near $T_{\rm C}$.
For bcc Fe and fcc Ni, we used the lattice constants obtained from structure optimization (Table~\ref{params}).
For reference, we also show literature $T_{\rm C}$ values computed via Monte Carlo simulations with an MFT-based method, using $J_{ij}$ parameters derived from FM and LMD states~\cite{Shallcross2005-ka}.
In Ref.~\cite{Shallcross2005-ka}, the magnitudes of the local magnetic moments in the LMD calculations were constrained to be equal to their ground-state values.}
\label{tc_elements}
\end{figure}

We also evaluated $T_{\rm C}$ using classical Monte Carlo simulations.
Figure~\ref{tc_elements} presents the systems directly comparable to experiment---bcc Fe, fcc Co, and fcc Ni---together with the experimental data~\cite{Haynes2016-rh}, while the complete results for all studied systems are summarized in Sec.~S10~\cite{SI}.
For fcc Co and fcc Ni, the (SC)$^2$ results for the adopted sampling regime appear consistent with experiment.
For bcc Fe, $T_{\rm C}$ estimated from the (SC)$^2$ method is somewhat higher than the experimental value; this is expected, because the present estimates neglect lattice-vibrational effects. 
Previous studies have shown that thermal atomic displacements substantially modify $J_{ij}$ in bcc Fe and thereby reduce $T_{\rm C}$~\cite{Ruban2018-mb, Mankovsky2020-em, Heine2021-qe, Tran2025-hf}.
In contrast, the corresponding vibrational corrections are reported to be much smaller for fcc Co and fcc Ni and, unlike in bcc Fe, do not lead to a substantial reduction of $T_{\rm C}$ (in some cases they even give a slight increase)~\cite{Ruban2018-mb, Mankovsky2020-em}.
Apart from that, our earlier work demonstrated that phonon softening associated with magnetic disorder thermodynamically stabilizes the paramagnetic state, further lowering $T_{\rm C}$~\cite{Tanaka2020-vf}.
For fcc Ni, however, phonon frequencies were shown to change only weakly between the FM and LMD states, suggesting a negligible impact on $T_{\rm C}$~\cite{Kormann2016-pn}, while for fcc Co systematic first-principles studies are, to the best of our knowledge, still lacking.
The magnitude of these phonon-related reductions for bcc Fe is summarized in Table~\ref{table:tc}.
Taken together---including the Nd-based magnet compounds---the (SC)$^2$ method yields physically reasonable magnetic phase transition temperatures once the sampling regime is fixed by the near-critical spin correlation.

\begin{table}
\caption{Summary of the effect of lattice vibrations on $T_{\rm C}$ in bcc Fe.
We distinguish two contributions here: (1) Variation of $T_{\rm C}$ through changes in $J_{ij}$ induced by thermal atomic displacements~\cite{Ruban2018-mb, Mankovsky2020-em, Heine2021-qe, Tran2025-hf}.
(2) Variation of $T_{\rm C}$ through phonon softening associated with magnetic disordering, which stabilizes the paramagnetic phase~\cite{Tanaka2020-vf}.
The referenced works evaluate $J_{ij}$ using MFT-based methods with the LMD state and estimate $T_{\rm C}$ via Monte Carlo simulations.}
\begin{tabular}{lrcc}
 	\hline \hline
	References & Change in $T_{\rm C}$ (K) \\
	\hline
	(1) Changes in $J_{ij}$ \\
	\quad Heine et al.~\cite{Heine2021-qe} & $\sim-400$  \\
	\quad Tran and Li~\cite{Tran2025-hf} & $-442$ \\
	\quad Ruban et al.~\cite{Ruban2018-mb} & $\sim-400$  \\
	\quad Mankovsky et al.~\cite{Mankovsky2020-em} & $\sim-230$  \\
	\hline 
	(2) Phonon softening\\
	\quad Tanaka and Gohda~\cite{Tanaka2020-vf} & $-560$ \\
	\hline 
\end{tabular}
\label{table:tc}
\end{table}

\section{Discussion}
\label{discussion}
\begin{figure}
\centering
\includegraphics[width=70mm]{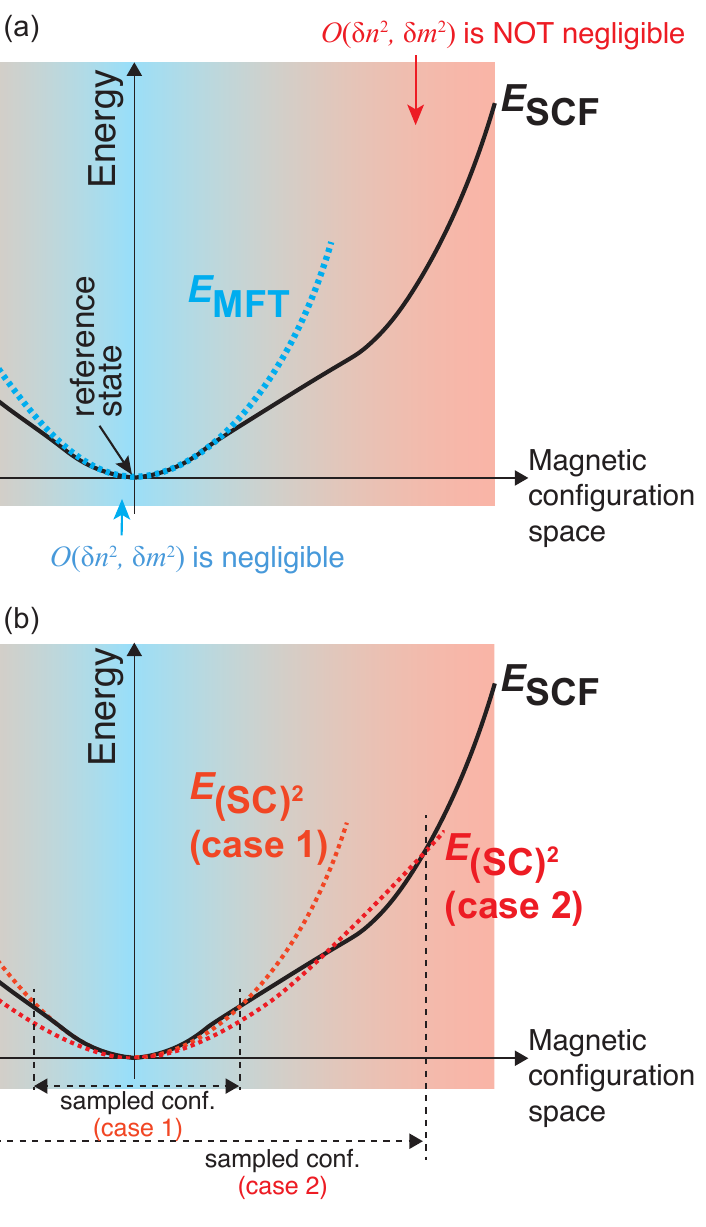}
\caption{Schematic representation of the energy landscape as a function of magnetic configuration around a reference magnetic state.
Black lines represent the energy obtained from self-consistent calculations ($E_{\rm SCF}$).
The blue dashed line in (a) represents the energy variation using $J_{ij}$ evaluated by the MFT-based method ($E_{\rm MFT}$).
The orange and red lines in (b) correspond to the energy variation using $J_{ij}$ evaluated by the (SC)$^2$ method ($E_{\rm (SC)^2}$) for different sampled magnetic configurations.
MFT-based methods are expected to reproduce the energy within the range where $\mathcal{O}(\delta n^2, \delta m^2)$ is negligible (indicated by the blue area).
In contrast, the (SC)$^2$ method constructs an effective model that reproduces the energy variation depending on the sampled magnetic configurations by considering not only $\delta^{*}E_{\rm band}$ but also $\mathcal{O}(\delta n^2, \delta m^2)$.
}
\label{schematic_energy}
\end{figure}

Throughout this study, we have examined the results from the viewpoint of electron--spin coupling.
Here we recast this viewpoint in the language of MFT, which provides a useful reference for delineating when a perturbative mapping is expected to remain quantitatively accurate. 

Within DFT, the energy variation $\delta E$ associated with infinitesimal spin rotations of magnetic moments can be decomposed as~\cite{Solovyev1998-jx, Katsnelson2004-pw}
\begin{align}
\begin{split}
\delta E &= \delta^{*}E_{\rm band} + \mathcal{O}(\delta n^2,\delta m^2), \\
&\approx \delta^{*}E_{\rm band},
\label{mft}
\end{split}
\end{align}
where $\delta^{*}E_{\rm band}$ denotes the band-energy variation evaluated at fixed charge density and fixed magnetization magnitude, and $\delta n$ and $\delta m$ represent changes in the charge density and the longitudinal component of the magnetization density, respectively.
The conventional MFT-based mapping corresponds to retaining only $\delta^{*}E_{\rm band}$.
For infinitesimal rotations, the remaining contribution $\mathcal{O}(\delta n^2,\delta m^2)$ is higher order and can be neglected.
Beyond this regime, however, finite rotations can induce spin-dependent rearrangements of the Kohn--Sham states, so that the associated self-consistent feedback on the energy becomes non-negligible.
In the present context, we refer to the energetic effect of this feedback as electron--spin coupling, and its contribution is represented by the $\mathcal{O}(\delta n^2,\delta m^2)$ term in Eq.~(\ref{mft}).
This perspective implies that, particularly for metallic systems where the occupation near the Fermi level can change under spin rotations, the range of rotation angles over which the MFT mapping remains quantitatively accurate can be narrower than is often assumed.

To illustrate this point, Fig.~\ref{schematic_energy} presents a schematic representation of the energy landscape as a function of magnetic configuration around a reference magnetic state.
The blue area in Fig.~\ref{schematic_energy}(a) represents a neighborhood in configuration space where the self-consistent feedback encoded in $\mathcal{O}(\delta n^2,\delta m^2)$ remains small, so that $E_{\rm MFT}$ can reproduce the self-consistent energy $E_{\rm SCF}$ (black line) with good accuracy.
Even when MFT-based approaches are refined or extended within an infinitesimal-rotation-based mapping, their quantitative accuracy is still expected to be highest in regimes where the fixed-density band-energy variation provides a good approximation to the total-energy change.
Accordingly, such approaches are naturally suited to the region indicated in blue, as illustrated by the blue dashed line in Fig.~\ref{schematic_energy}(a).

Outside this range, there inevitably exists a region in which the self-consistent electronic response must be taken into account to reproduce $E_{\rm SCF}$ quantitatively.
In this region, $E_{\rm SCF}$ can exhibit complex variations driven by electron--spin coupling, as demonstrated in SrMnO$_3$ at $a/a_{\rm eq}=1.05$.
Even under such conditions, a Heisenberg model can still serve as a useful phenomenological description provided that it sufficiently reproduces $E_{\rm SCF}$ over the targeted portion of configuration space.
However, because the contribution of electron--spin coupling depends on the sampled magnetic configurations, the fitted parameters $J_{ij}$ can acquire a dependence on the sampling window.
This is illustrated in Fig.~\ref{schematic_energy}(b), where different sampling choices (case 1 and case 2) lead to different effective $E_{\rm (SC)^2}$ curves and therefore different effective $J_{ij}$.
In this sense, the configuration dependence of the extracted $J_{ij}$ is not a numerical artifact but rather a quantitative signature of electron--spin coupling in the underlying first-principles energy landscape.

\section{Conclusion}
\label{summary}
We have shown that finite spin rotations can induce self-consistent changes in the electronic structure that feed back on the effective exchange couplings $J_{ij}$, i.e., electron--spin coupling.
In the language of MFT, this feedback contributes to the energy terms beyond the band-energy variation evaluated at a frozen density and fixed moment magnitude, and it can therefore modify the $J_{ij}$ extracted from finite-angle configurations.
For SrMnO$_3$ with the type--G antiferromagnetic state, the previously reported discrepancy between conventional MFT-based evaluations and self-consistent finite-rotation results~\cite{Zhu2020-ik} can be understood in terms of spin-rotation-induced spin splitting and the associated changes in band occupancy, which amplify the above electronic feedback.
While the conventional MFT mapping correctly captures the energy curvature in the strict infinitesimal-rotation limit, describing finite rotations and the resulting energy landscape requires a fully self-consistent, nonperturbative treatment.
For the Nd-based permanent-magnet compounds, our nonperturbative analysis reveals a pronounced difference in the sensitivity of $J_{ij}$ and $T_{\rm C}$ to magnetic disorder between Nd$_2$Fe$_{14}$B and Nd$_2$Co$_{14}$B:
Nd$_2$Fe$_{14}$B exhibits a rapid decrease in $T_{\rm C}$ with increasing magnetic disorder, whereas Nd$_2$Co$_{14}$B shows a much weaker dependence.
These results provide insight into the quantitative composition dependence of $J_{ij}$ and $T_{\rm C}$ in Nd-based permanent magnets.
For elemental 3$d$ transition metals with bcc and fcc structures, we observe substantial discrepancies in the first-nearest-neighbor $J_{ij}$ relative to a conventional MFT-based evaluation.
The pronounced angular dependence of the effective $J_{ij}$ in our self-consistent sampling suggests that electron--spin coupling can provide an important contribution even in these comparatively simple metallic systems.

The self-consistent supercell calculations referred to as the (SC)$^2$ method---conceptually simple and now computationally practical with current resources---yield internally consistent $J_{ij}$ across a wide angular range, from small spin rotations to substantially larger magnetic disorder, thereby bridging the gap between the infinitesimal- and finite-angle regimes.
In this sense, the (SC)$^2$ method and MFT-based perturbative approaches are complementary and mutually reinforcing; perturbative methods provide efficient and interpretable descriptions such as orbital-resolved analyses~\cite{Korotin2015-eg}, whereas the nonperturbative scheme explicitly captures the electron--spin coupling, which becomes relevant beyond the infinitesimal limit.
A transferable protocol for choosing the sampling regime, for example by determining it self-consistently with the resulting spin model, remains a necessary step toward quantitative spin-model construction.
Advancing both the perturbative and nonperturbative approaches will be essential for the analysis and design of magnetic materials.
We plan to further develop the nonperturbative method and release it to the community in a user-friendly format.

\begin{acknowledgments}
This work was partly supported by JSPS KAKENHI Grant Number JP24K01144 and MEXT-DXMag Grant Number JPMXP1122715503. The calculations were partly carried out by using facilities of the Supercomputer Center at the Institute for Solid State Physics, the University of Tokyo, and TSUBAME4.0 supercomputer at Institute of Science Tokyo.
\end{acknowledgments}

\section*{Conflict of Interest}
The authors declare no conflicts of interest.

\end{document}


\title{Supplementary Information \\
Impact of electron--spin coupling on exchange coupling parameters:\\a nonperturbative approach}
\author{Tomonori Tanaka}
\author{Yoshihiro Gohda}
\affiliation{Department of Materials Science and Engineering, Institute of Science Tokyo, Yokohama 226-8501, Japan}
\date{\today}%
\pacs{}%
\maketitle
\tableofcontents

\clearpage

\section{$J_{ij}$ of SrMnO$_3$}
\begin{figure}[h]
\centering
\includegraphics[width=\linewidth]{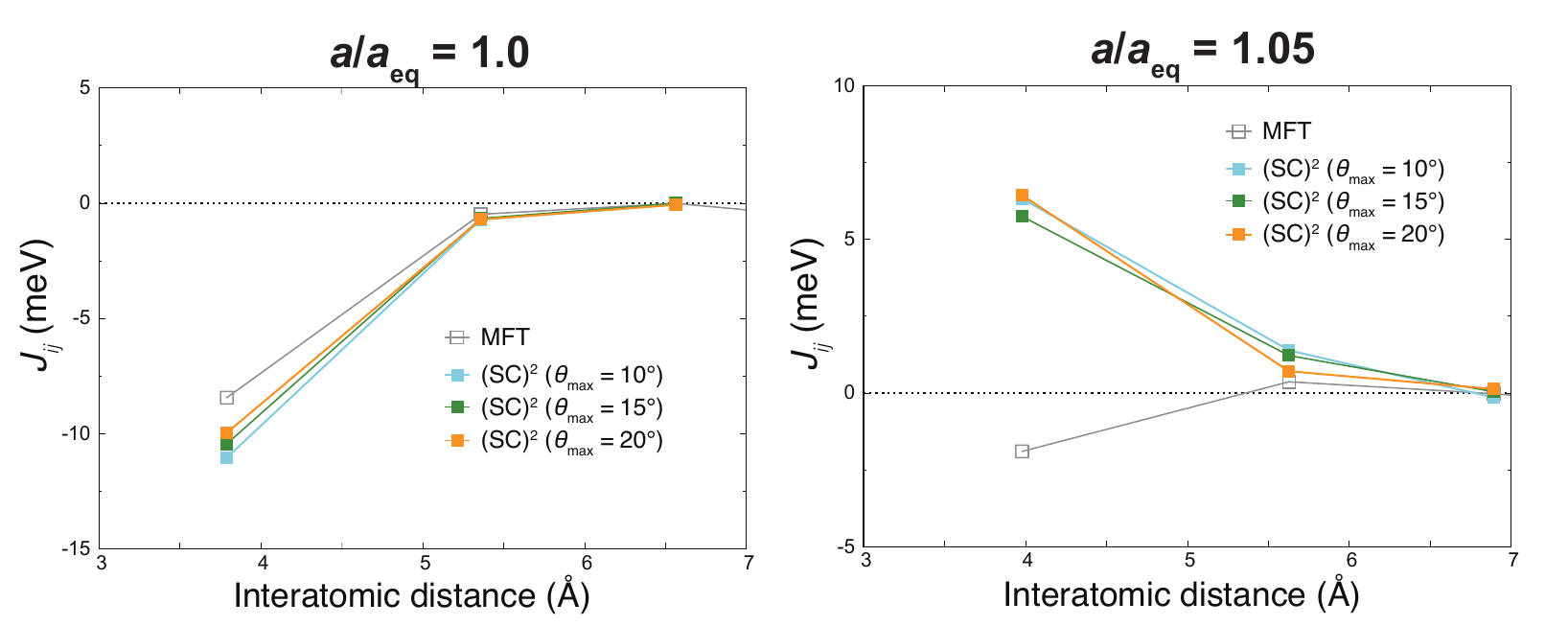}
\caption{
Exchange coupling parameters ($J_{ij}$) between Mn atoms in cubic perovskite SrMnO$_3$ with a type--G antiferromagnetic state.
The gray data correspond to the MFT-based method, while the blue, green, and orange data represent the (SC)$^2$ method results obtained with upper bounds of 10$^\circ$, 15$^\circ$, and 20$^\circ$, respectively.
}
\label{srmno3_jij}
\end{figure}

In the main text, we presented only $J_{ij}$ for the first-nearest-neighbor Mn--Mn pairs ($J_{01}$).
For reference, Fig.~\ref{srmno3_jij} shows the $J_{ij}$ values for more distant pairs.

\begin{figure}[h]
\centering
\includegraphics[width=90mm]{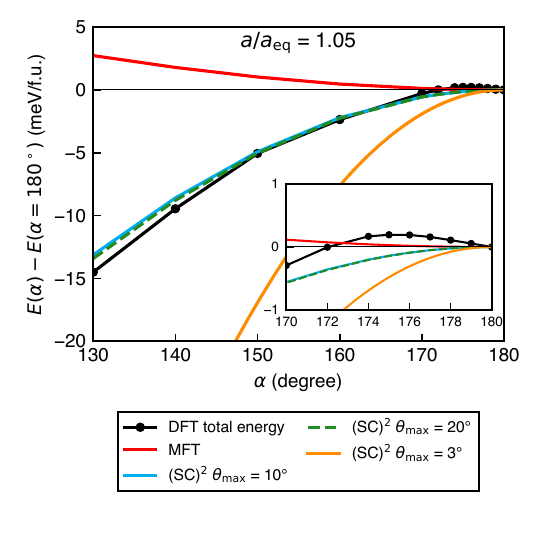}
\caption{
Total energy variation per formula unit of SrMnO$_3$ at $a/a_{\rm eq}=1.05$ as a function of the rotation angle $\alpha$ of the Mn magnetic moments, as in Fig.~2 of the main text, with the Heisenberg curve from the (SC)$^2$ fit for $\theta_{\rm max}=3^\circ$ added (orange).
The inset shows a magnified view between $170^\circ$ and $180^\circ$.
}
\label{srmno3_theta3}
\end{figure}

We also examined whether the local minimum at $\alpha=180^\circ$ in Fig.~2 of the main text ($a/a_{\rm eq}=1.05$) is recovered when the sampled configurations are restricted to its immediate neighborhood, by performing the (SC)$^2$ fit with $\theta_{\rm max}=3^\circ$ and all other settings the same as for $\theta_{\rm max}=10^\circ$ and $20^\circ$.
As shown in Fig.~\ref{srmno3_theta3}, the local minimum is not recovered, and the fitted curve deviates further from the self-consistent total energies near $\alpha=180^\circ$.
Although the energy differences involved are small (the energies of the 28 sampled configurations span about 1.5 meV per $2\times2\times2$ supercell), they are well resolved: the self-consistent total energies are converged to better than $10^{-7}$ eV per atom, and the root mean squared error of the Heisenberg fit is 0.04 meV per supercell.
The fitted $J_{ij}$ for $\theta_{\rm max}=3^\circ$ differ markedly from those for $10^\circ$ and $20^\circ$ and do not describe the local curvature of the energy landscape at this gapless state.
This is because the energies of the randomly canted configurations, in which the Mn moments tilt independently, lie both above and below the energy of the type--G antiferromagnetic state even for canting angles within a few degrees (between $-0.13$ and $+0.06$ meV per formula unit; not shown), so that the local minimum along the coherent rotation of the two sublattices in Fig.~2 is absent from the sampled energies.
The energy change thus depends on the canting pattern and not only on the relative angles between the moments.

\clearpage
\section{Magnetic moments in Nd$_2$Fe$_{14}$B and Nd$_2$Co$_{14}$B}
In the Nd$_2$Fe$_{14}$B and Nd$_2$Co$_{14}$B systems, we used two different DFT packages: VASP for the (SC)$^2$ method and AkaiKKR for the MFT-based method.
To verify that the results obtained from these two packages are sufficiently consistent, we compare the converged magnetic moments in the ground states, as summarized in Table~\ref{magmom}.
The results show that both packages yield magnetic states that are in reasonable agreement.

\begin{table}[h]
\caption{Magnetic moments $\mathfrak{m}$ ($\mu_{\rm B}$) and atomic radii (\AA) used in the Nd$_2$Fe$_{14}$B and Nd$_2$Co$_{14}$B systems. The radii in the AkaiKKR package represent the muffin-tin radii, whereas those in VASP correspond to the RWIGS parameters. Note that the magnetic moments of Nd do not include the contribution of the 4$f$ electrons.}
\begin{tabular}{lrcc}
 	\hline \hline
	Wyckoff position & $\mathfrak{m}$ (radius) in AkaiKKR & \quad  $\mathfrak{m}$ (radius) in VASP\\
	\hline
	Nd$_2$Fe$_{14}$B \\
	\quad Nd $4f$ & $-$0.40 (1.80) & $-$0.25 (1.50) \\
	\quad Nd $4g$ & $-$0.28 (1.76) & $-$0.26 (1.50)\\
	\quad Fe $4c$ & 2.53 (1.29) & 2.50 (1.20)\\
	\quad Fe $4e$ & 1.80 (1.10) & 2.06 (1.20)\\
	\quad Fe $8j_1$ & 2.43 (1.19) & 2.32 (1.20)\\
	\quad Fe $8j_2$ & 2.68 (1.23) & 2.74 (1.20)\\
	\quad Fe $16k_1$ & 2.11 (1.13) & 2.29 (1.20)\\
	\quad Fe $16k_2$ & 2.27 (1.19) & 2.39 (1.20)\\
	\quad B $4g$ & $-$0.18 (0.96) & $-$0.11 (0.80)\\
	\hline 
	Nd$_2$Co$_{14}$B \\
	\quad Nd $4f$ & $-$0.31 (1.78) & $-$0.17 (1.50) \\
	\quad Nd $4g$ & $-$0.24 (1.77) & $-$0.17 (1.50) \\
	\quad Co $4c$ & 1.77 (1.28) & 1.61 (1.20) \\
	\quad Co $4e$ & 0.60 (1.07) & 0.95 (1.20) \\
	\quad Co $8j_1$ & 1.73 (1.17) & 1.53 (1.20) \\
	\quad Co $8j_2$ & 1.66 (1.20) & 1.63 (1.20) \\
	\quad Co $16k_1$ & 1.08 (1.11) & 1.23 (1.20)\\
	\quad Co $16k_2$ & 1.44 (1.17) & 1.47 (1.20)\\
	\quad B $4g$ & $-$0.09 (0.96) & $-$0.06 (0.80)\\
	\hline 
\end{tabular}
\label{magmom}
\end{table}

\clearpage
\section{Curie temperatures of Nd$_2$Fe$_{14}$B and Nd$_2$Co$_{14}$B}
In the main text, we showed the difference in the Curie temperatures between Nd$_2$Fe$_{14}$B and Nd$_2$Co$_{14}$B.
Here, we present the absolute values of the Curie temperatures of Nd$_2$Fe$_{14}$B and Nd$_2$Co$_{14}$B obtained from experiments and theoretical calculations.
The theoretical values are obtained by the mean-field approximation~\cite{Matsubara1977-jf}.
We note that the overestimation of the Curie temperatures in the (SC)$^2$ method reflects an expected trend for several reasons: the mean-field approximation systematically overestimates phase transition temperatures, and recent studies have revealed that certain interactions between magnetism and phonons also contribute to lowering magnetic phase transition temperatures~\cite{Heine2021-qe, Tsuna2023-fa, Tanaka2020-vf}.

\begin{figure}[h]
\centering
\includegraphics[width=130mm]{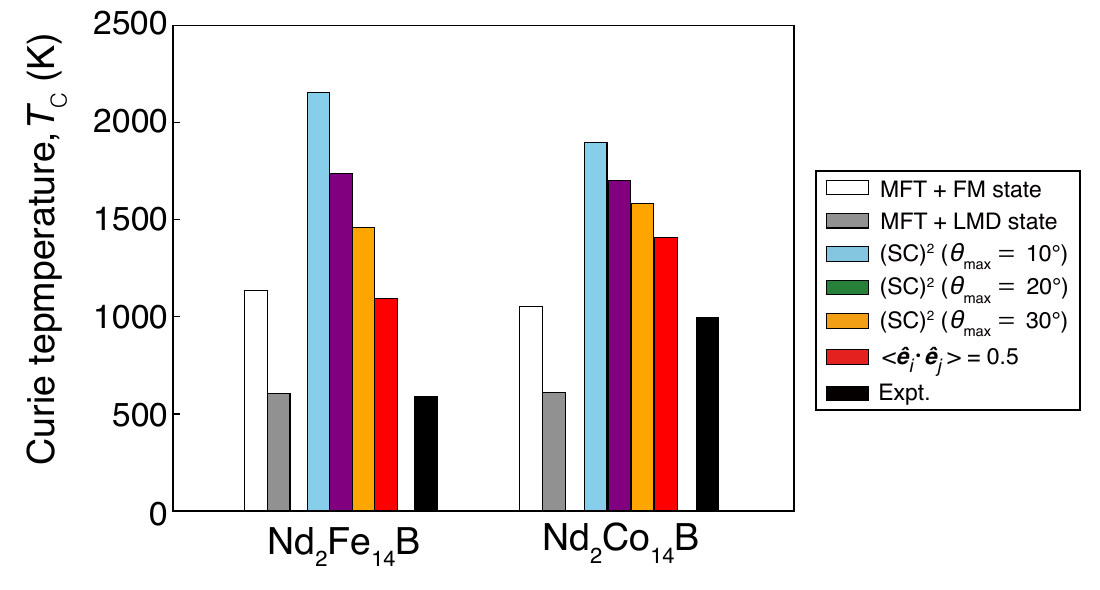}
\caption{
Curie temperatures ($T_{\rm C}$) of Nd$_2$Co$_{14}$B and Nd$_2$Fe$_{14}$B.
The white and gray bars show values obtained with the MFT-based method using ferromagnetic (FM) and local moment disorder (LMD) reference states, respectively.
The black bars represent the experimental values~\cite{Sagawa1984-rp, Fuerst1988-qm}.
The blue, green, and orange bars indicate values obtained by the (SC)$^2$ method, with the upper bound of $\theta$ set to $10^\circ$, $20^\circ$, and $30^\circ$, respectively.
The red bars indicate the values obtained by the (SC)$^2$ method using the spin-direction distribution in Eq.~(\ref{prob}) with $\tau=0.6261$, chosen to reproduce the nearest-neighbor spin--spin correlation at the Curie temperature of bcc Fe reported in Refs.~\cite{Melnikov2019-es, Walsh2022-ci}, $\langle \hat{\bm e}_i \cdot \hat{\bm e}_j \rangle = 0.5$.
Theoretical values are obtained by the mean-field approximation~\cite{Matsubara1977-jf}.
}
\label{tc_2-14-1}
\end{figure}

\clearpage
\section{Sensitivity of $J_{ij}$ to atomic sphere radius}
\begin{figure}[h]
\centering
\includegraphics[width=\linewidth]{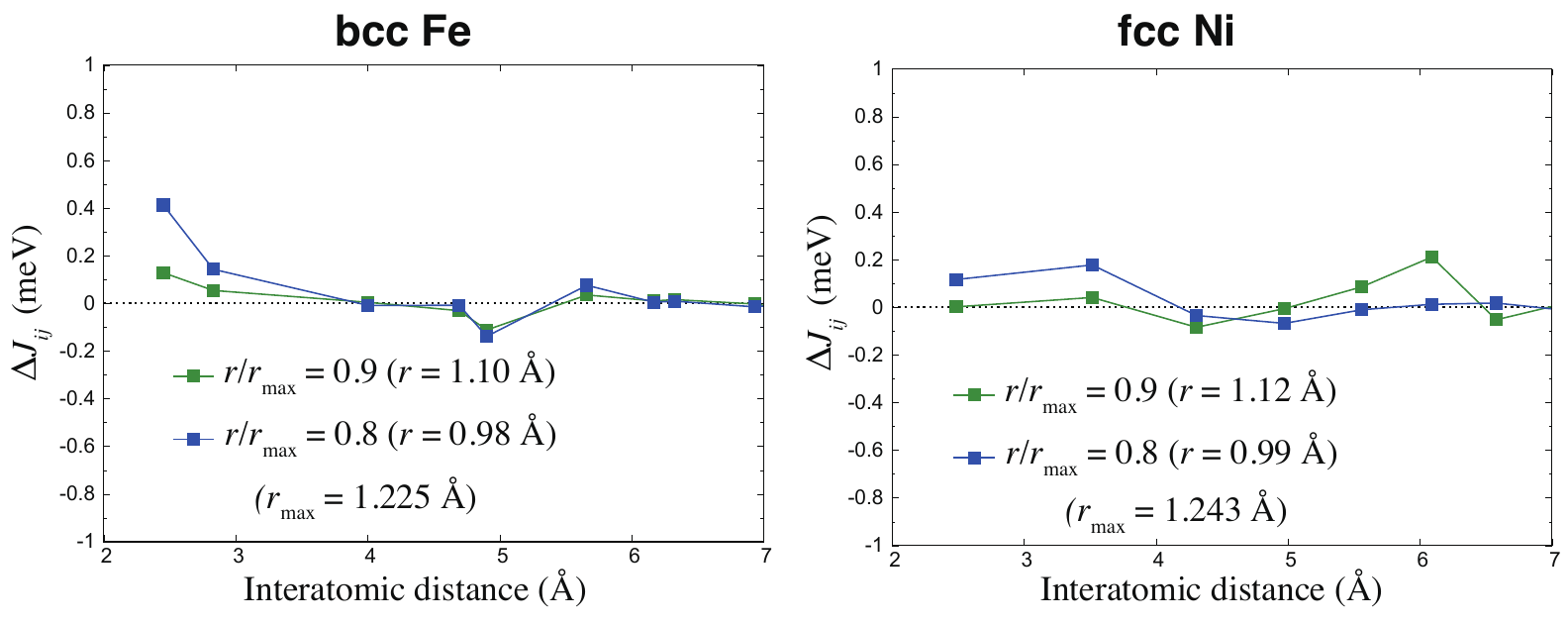}
\caption{
Dependence of $J_{ij}$ on atomic sphere radius used for applying the external magnetic field for bcc Fe (left panel) and fcc Ni (right panel).
}
\label{rwigs}
\end{figure}

We checked the sensitivity of $J_{ij}$ to the atomic sphere radius (VASP parameter RWIGS), which is a parameter in the constrained noncollinear spin DFT calculations.
Figure~\ref{rwigs} shows the differences in $J_{ij}$ relative to those obtained when the RWIGS parameter ($r$) is set to its maximum value ($r_{\rm max}$), defined as
\begin{align}
\Delta J_{ij} = J^{r/r_{\rm max} = 0.9\ {\rm or}\ 0.8}_{ij} - J^{r/r_{\rm max} = 1} _{ij}. \nonumber
\end{align}
From the results, we reasonably conclude that the sensitivity of $J_{ij}$ to the atomic sphere radius is negligibly small.
Note that the results in the main text are those obtained when $r/r_{\rm max} = 1$.

\clearpage
\section{Dependence of $J_{ij}$ on supercell size}
Finite-size effects due to interactions between periodic images are inevitable in the supercell method.
Consequently, the obtained $J_{ij}$ values reflect the renormalized interactions of distant and equivalent atomic pairs. 
To ensure quantitative reliability, we checked the convergence of $J_{ij}$ values with respect to supercell sizes for elemental 3$d$ transition metals.

\begin{figure}[h]
\centering
\includegraphics[width=\linewidth]{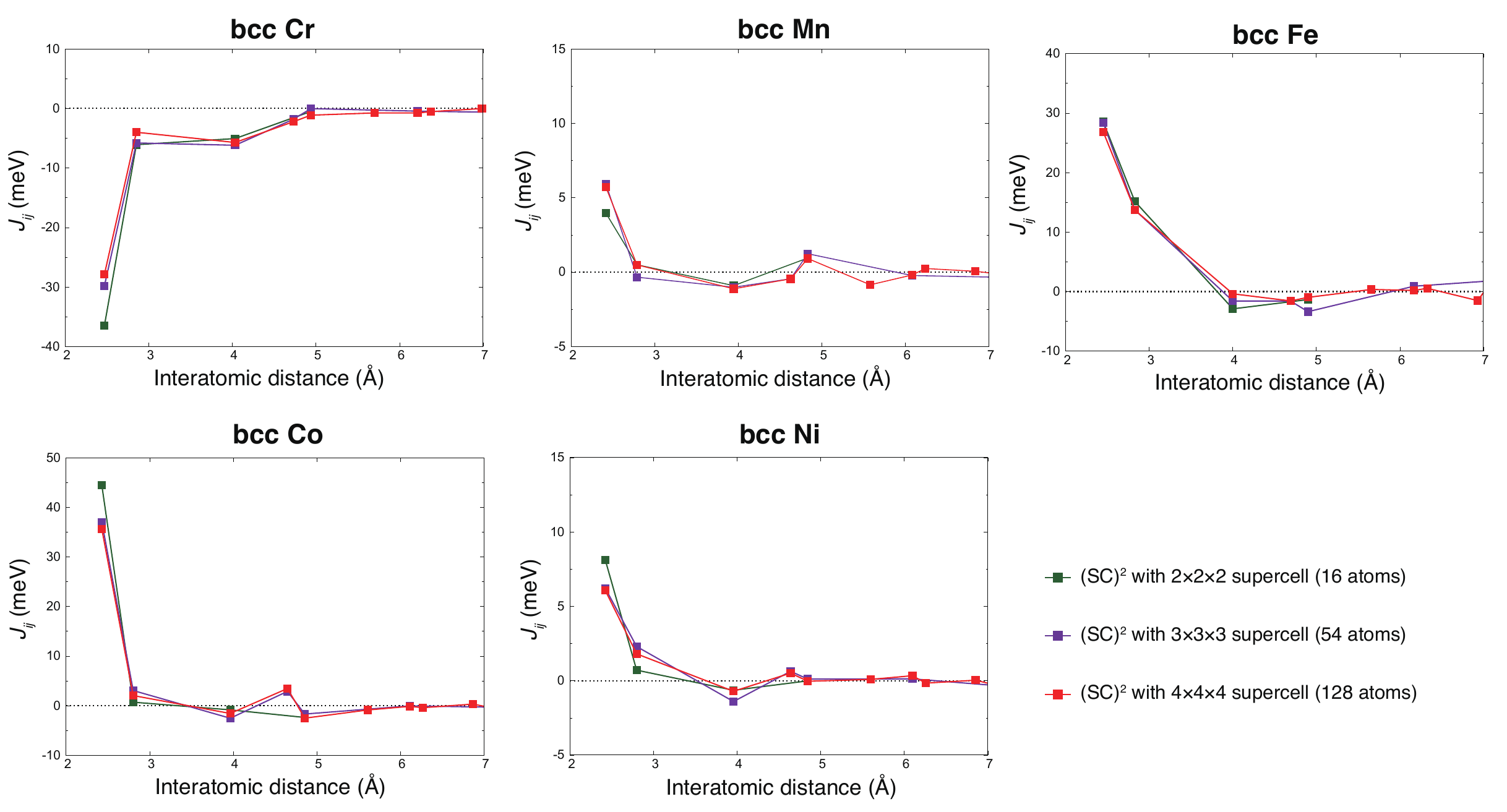}
\caption{
Dependence of $J_{ij}$ on supercell size for bcc systems.
The green, purple, and red lines represent the different supercell sizes of $2\times2\times2$, $3\times3\times3$, and $4\times4\times4$, respectively.
}
\label{jijbcc_size}
\end{figure}

\begin{figure}[h]
\centering
\includegraphics[width=\linewidth]{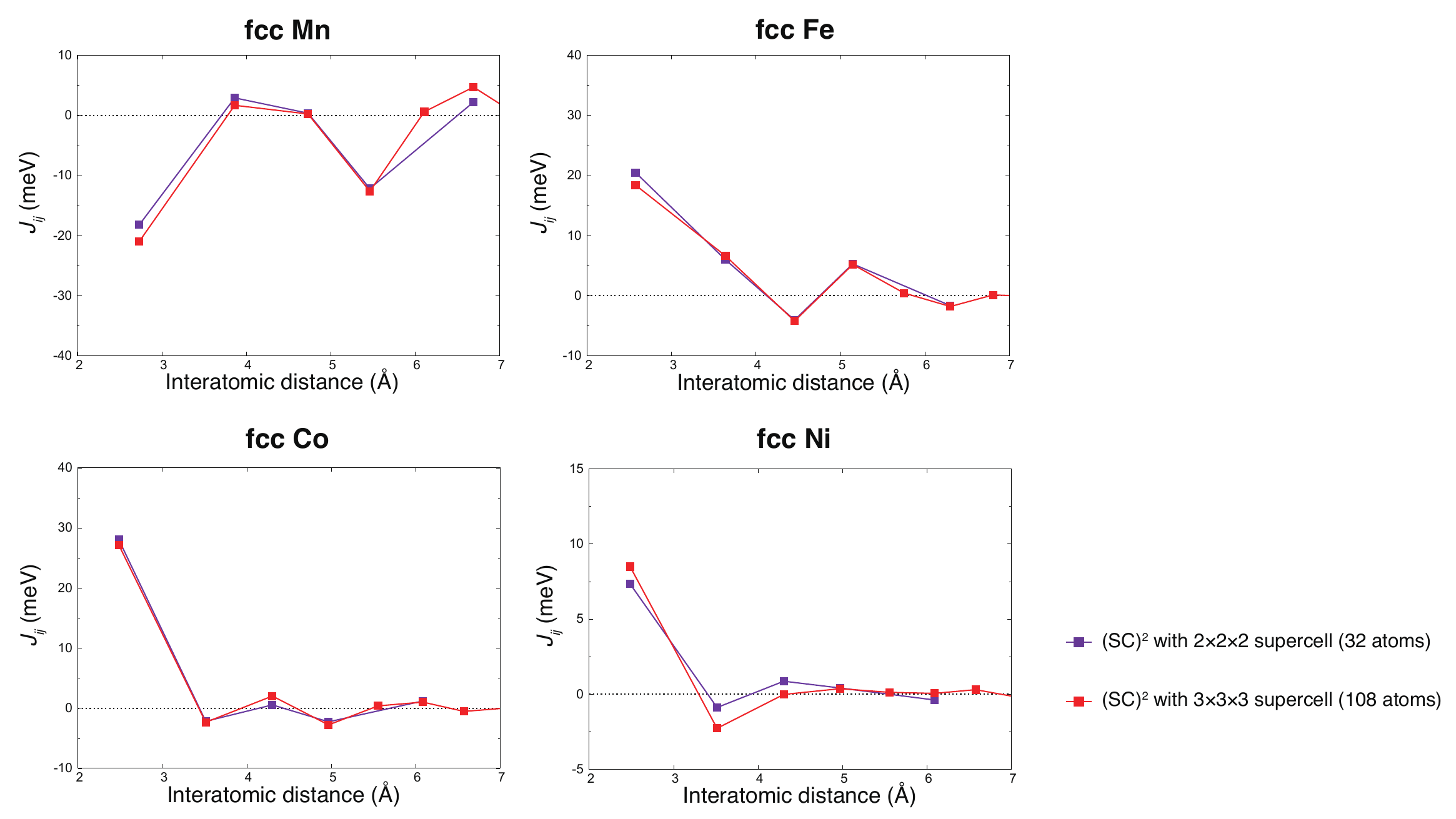}
\caption{
Dependence of $J_{ij}$ on supercell size for fcc systems.
The purple and red lines represent the different supercell sizes of $2\times2\times2$ and $3\times3\times3$, respectively.
}
\label{jijfcc_size}
\end{figure}

Figures~\ref{jijbcc_size} and \ref{jijfcc_size} show $J_{ij}$ for different supercell sizes.
Based on these results, a $4\times4\times4$ supercell for bcc and a $3\times3\times3$ supercell for fcc achieve good convergence.

\clearpage
\section{Band structures of the tight-binding model using Wannier functions}
\begin{figure}[h]
\centering
\includegraphics[width=\linewidth]{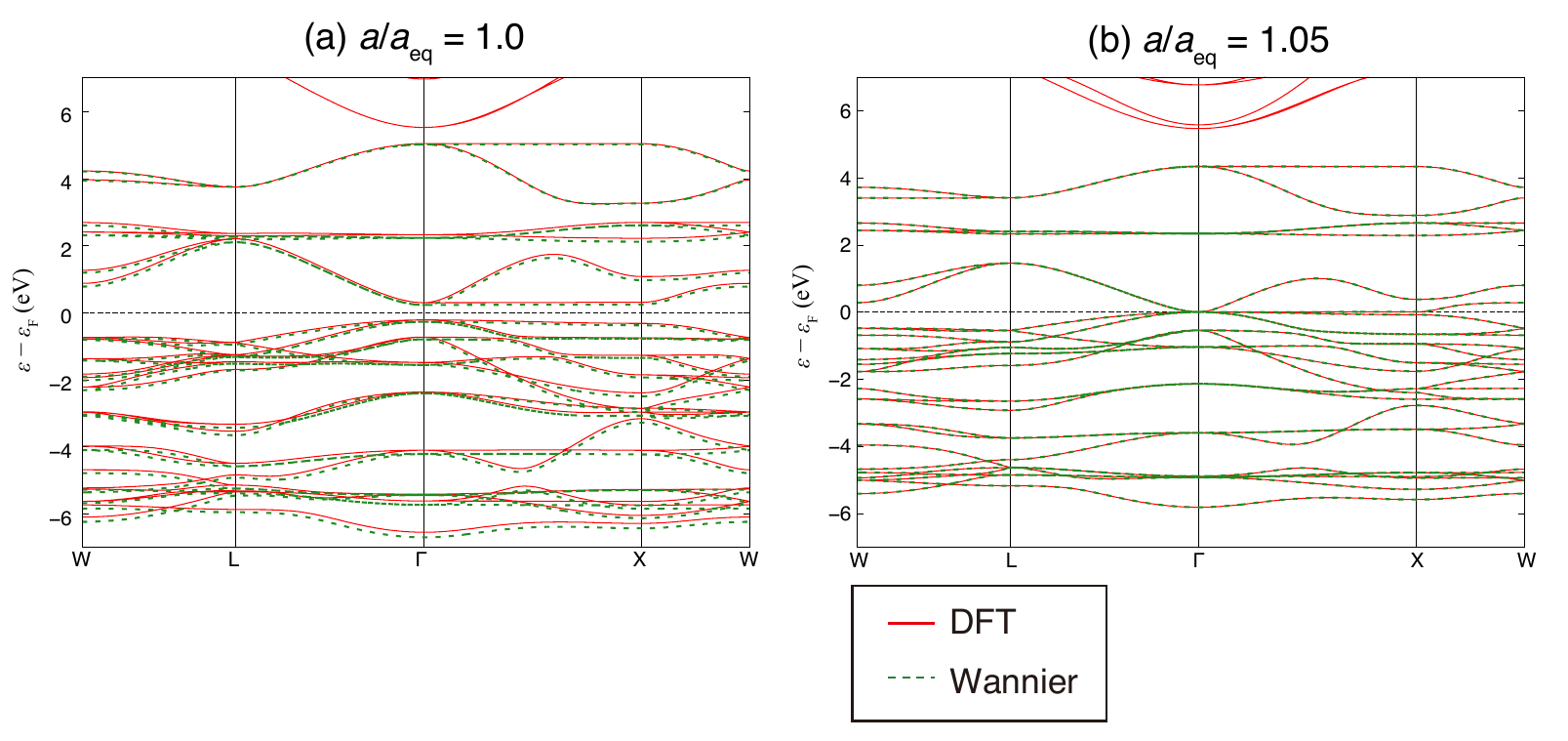}
\caption{Band structure of SrMnO$_3$ in the type--G antiferromagnetic state.
Red lines and green dashed lines are those calculated by DFT and tight-binding calculations using Wannier functions.
Note that the up and down bands are completely overlapped.}
\label{wannier_srmno3}
\end{figure}

\begin{figure}[h]
\centering
\includegraphics[width=140mm]{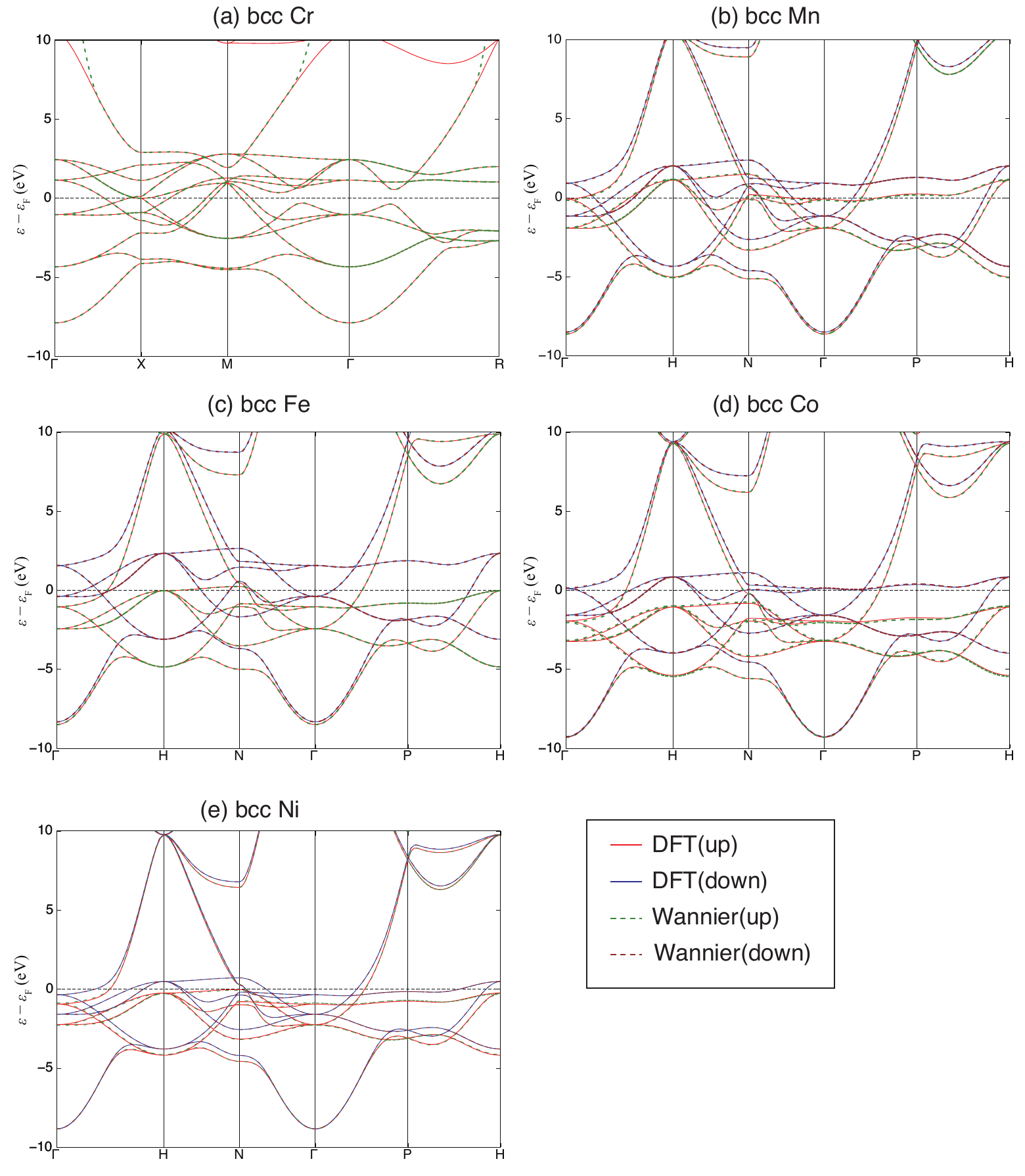}
\caption{Band structure of (a) bcc Cr, (b) bcc Mn, (c) bcc Fe, (d) bcc Co, and (e) bcc Ni.
Red and blue lines are those for the up and down spins calculated by DFT and green and brown dashed lines are those by tight-binding calculations using Wannier functions.}
\label{wannier_bcc}
\end{figure}

\begin{figure}[h]
\centering
\includegraphics[width=140mm]{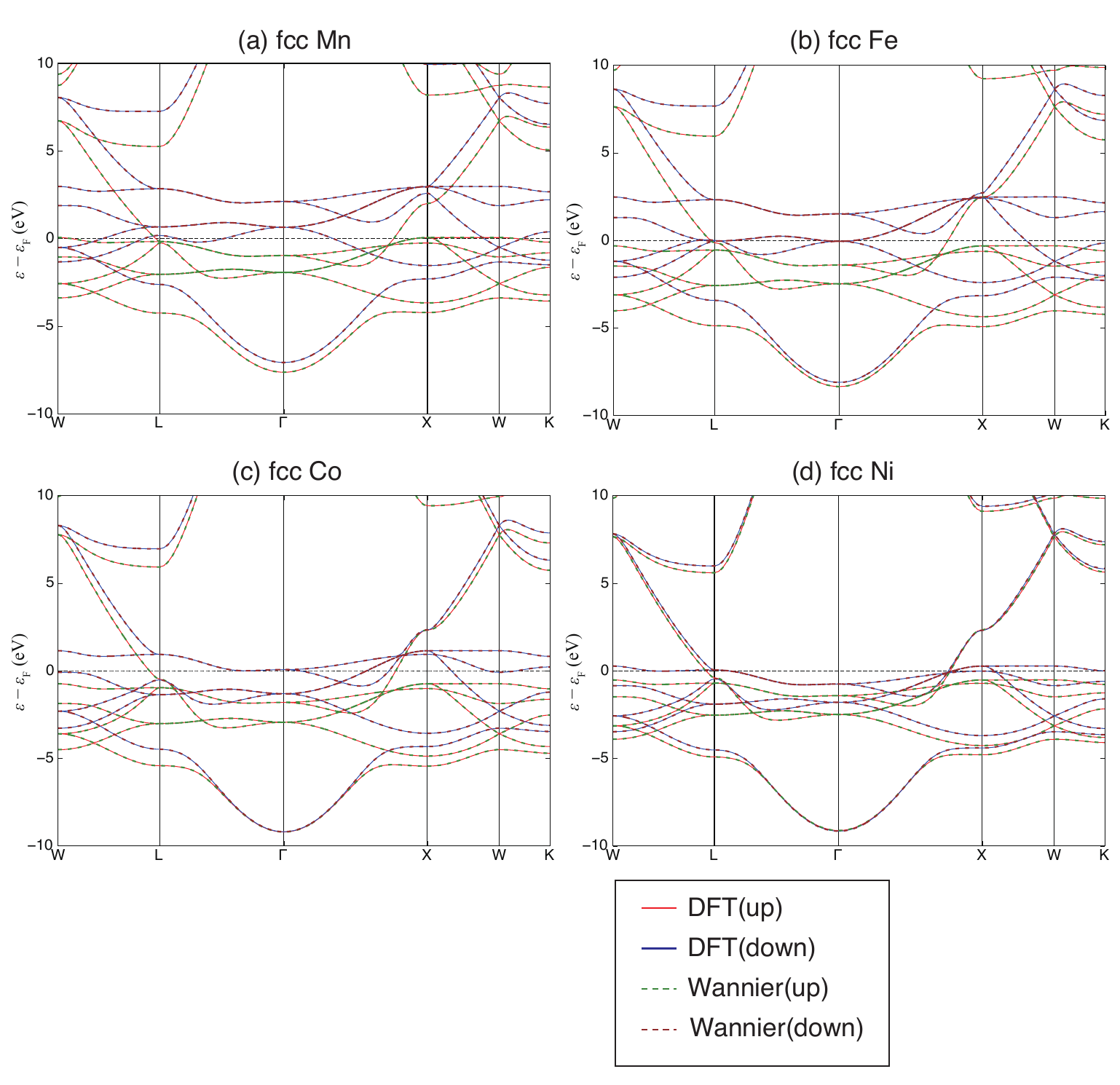}
\caption{Band structure of (a) fcc Mn, (b) fcc Fe, (c) fcc Co, and (d) fcc Ni. All of the figures were plotted in the same manner as Fig.~\ref{wannier_bcc}.}
\label{wannier_fcc}
\end{figure}

To demonstrate that the constructed Wannier functions possess sufficient quality, we compare band structures obtained from DFT and tight-binding calculations using the Wannier functions as shown in Figs.~\ref{wannier_srmno3}, \ref{wannier_bcc}, and \ref{wannier_fcc} for SrMnO$_3$, bcc elements, and fcc elements, respectively.
For the projected angular momentum states, we used the $d$ orbitals for Mn and the $p$ orbitals for O in SrMnO$_3$, whereas $s$, $p$, and $d$ orbitals were used for the bcc and fcc elements.

\clearpage
\section{Derivation of $\langle(\hat{{\bm e}}_{i} \cdot \hat{\bm e}_{j})^2\rangle_{\theta_{\rm max}}/\langle\hat{{\bm e}}_{i} \cdot \hat{\bm e}_{j}\rangle_{\theta_{\rm max}}$}
Here, we derive the analytical form of $\langle(\hat{{\bm e}}_{i} \cdot \hat{\bm e}_{j})^2\rangle_{\theta_{\rm max}}/\langle\hat{{\bm e}}_{i} \cdot \hat{\bm e}_{j}\rangle_{\theta_{\rm max}}$.
A unit vector $\hat{{\bm e}}_{i}$ is defined as
\begin{align}
\hat{{\bm e}}_{i} = \sin\theta_i\cos\phi_i{\bm{\hat{x}}} +  \sin\theta_i\sin\phi_i{\bm{\hat{y}}} + \cos\theta_i{\bm{\hat{z}}},\nonumber
\end{align}
therefore
\begin{align}
\hat{{\bm e}}_{i} \cdot \hat{\bm e}_{j}
&= \sin\theta_i\sin\theta_j\cos\phi_i\cos\phi_j + \sin\theta_i\sin\theta_j\sin\phi_i\sin\phi_j
+\cos\theta_i\cos\theta_j \nonumber \\
&= \sin\theta_i\sin\theta_j\cos(\phi_i - \phi_j) + \cos\theta_i\cos\theta_j, \nonumber
\end{align}

\begin{align}
(\hat{{\bm e}}_{i} \cdot \hat{\bm e}_{j})^2 \nonumber
=&\sin^2\theta_i\sin^2\theta_j\cos^2(\phi_i - \phi_j) + 2\sin\theta_i\sin\theta_j\cos\theta_i\cos\theta_j\cos(\phi_i - \phi_j)  \\
&+ \cos^2\theta_i\cos^2\theta_j. \nonumber
\end{align}

The expectations with $\theta_{\rm max}$ are
\begin{align}
\langle\hat{{\bm e}}_{i} \cdot \hat{\bm e}_{j} \rangle_{\theta_{\rm max}}
&= \frac{1}{(2\pi(1-\cos\theta_{\rm max}))^2} \times \nonumber \\
&\ \int_{0}^{\theta_{\rm max}}\int_{0}^{\theta_{\rm max}}\int_{0}^{2\pi}\int_{0}^{2\pi} \hat{{\bm e}}_{i} \cdot \hat{{\bm e}}_{j} \sin\theta_i\sin\theta_j d\phi_id\phi_jd\theta_id\theta_j \nonumber\\
&= \frac{4\pi^2}{4\pi^2(1-\cos\theta_{\rm max})^2} \int_{0}^{\theta_{\rm max}}\int_{0}^{\theta_{\rm max}}
\cos\theta_i\cos\theta_j \sin\theta_i\sin\theta_j d\theta_id\theta_j \nonumber \\
&= \frac{1}{(1-\cos\theta_{\rm max})^2}\left( \frac{1}{2}\sin^2\theta_{\rm max}\right)^2 \nonumber \\
&=\frac{\sin^4\theta_{\rm max}}{4(1-\cos\theta_{\rm max})^2},\label{corr_uni}
\end{align}
and
\begin{align}
\langle(\hat{{\bm e}}_{i} \cdot \hat{\bm e}_{j})^2 \rangle_{\theta_{\rm max}}  \nonumber
&= \frac{1}{(2\pi(1-\cos\theta_{\rm max}))^2} \times \nonumber\\
&\ \int_{0}^{\theta_{\rm max}}\int_{0}^{\theta_{\rm max}}\int_{0}^{2\pi}\int_{0}^{2\pi} (\hat{{\bm e}}_{i} \cdot \hat{{\bm e}}_{j})^2 \sin\theta_i\sin\theta_j d\phi_id\phi_jd\theta_id\theta_j \nonumber\\
&= \frac{1}{(2\pi(1-\cos\theta_{\rm max}))^2} \times \nonumber\\
&\int_{0}^{\theta_{\rm max}}\int_{0}^{\theta_{\rm max}}\left(2\pi^2\sin^3\theta_i\sin^3\theta_j
+ 4\pi^2\cos^2\theta_i\cos^2\theta_j\sin\theta_i\sin\theta_j\right) d\theta_id\theta_j \nonumber\\
&= \frac{2}{4(1-\cos\theta_{\rm max})^2}\left(\left(-\frac{3}{4}\cos\theta_{\rm max}+\frac{1}{12}\cos3\theta_{\rm max} + \frac{2}{3}\right)^2 + \frac{2}{9}\left(1-\cos^3\theta_{\rm max}\right)^2\right),
\end{align}
where $2\pi(1-\cos\theta_{\rm max})$ is the area of the spherical cap with $\theta_{\rm max}$.
Finally, we obtain
\begin{align}
\frac{\langle(\hat{{\bm e}}_{i} \cdot \hat{\bm e}_{j})^2 \rangle_{\theta_{\rm max}}}{\langle\hat{{\bm e}}_{i} \cdot \hat{\bm e}_{j} \rangle_{\theta_{\rm max}}}
= \frac{2}{\sin^4\theta_{\rm max}}\left(\left(-\frac{3}{4}\cos\theta_{\rm max}+\frac{1}{12}\cos3\theta_{\rm max} + \frac{2}{3}\right)^2 + \frac{2}{9}\left(1-\cos^3\theta_{\rm max}\right)^2\right).
\end{align}

\clearpage
\section{Performance of the Heisenberg model as a regression model}
\begin{figure}[h]
\centering
\includegraphics[width=120mm]{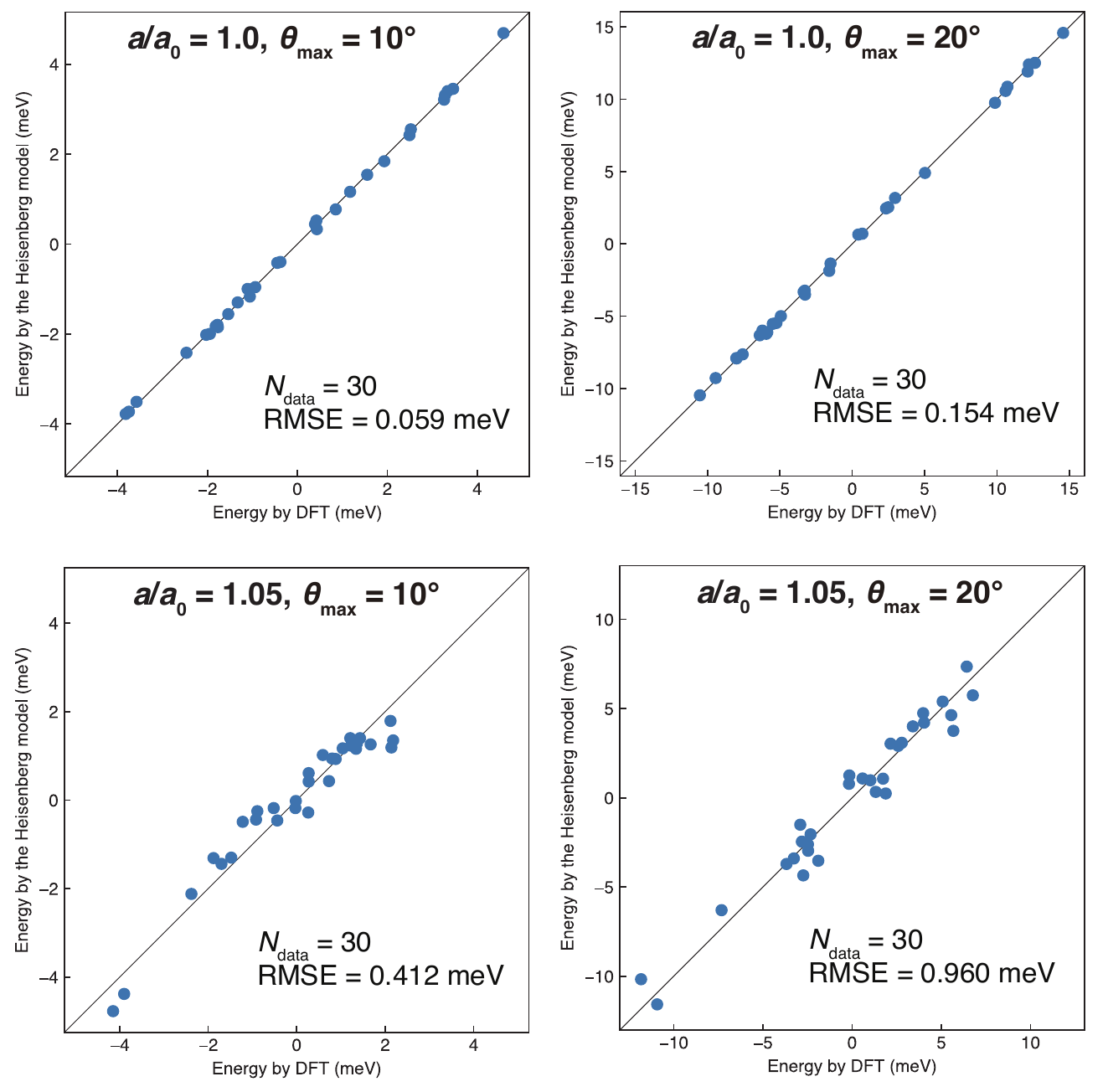}
\caption{Comparison between the energies calculated using DFT and the Heisenberg model for SrMnO$_3$ in the type--G antiferromagnetic state.
The RMSE and the total number of magnetic configurations $N_{\rm data}$ are also provided for each figure.
Note that the origin of the energies is defined as the mean value of the DFT energy dataset.
 }
\label{energy_srmno3}
\end{figure}

\begin{figure}[h]
\centering
\includegraphics[width=120mm]{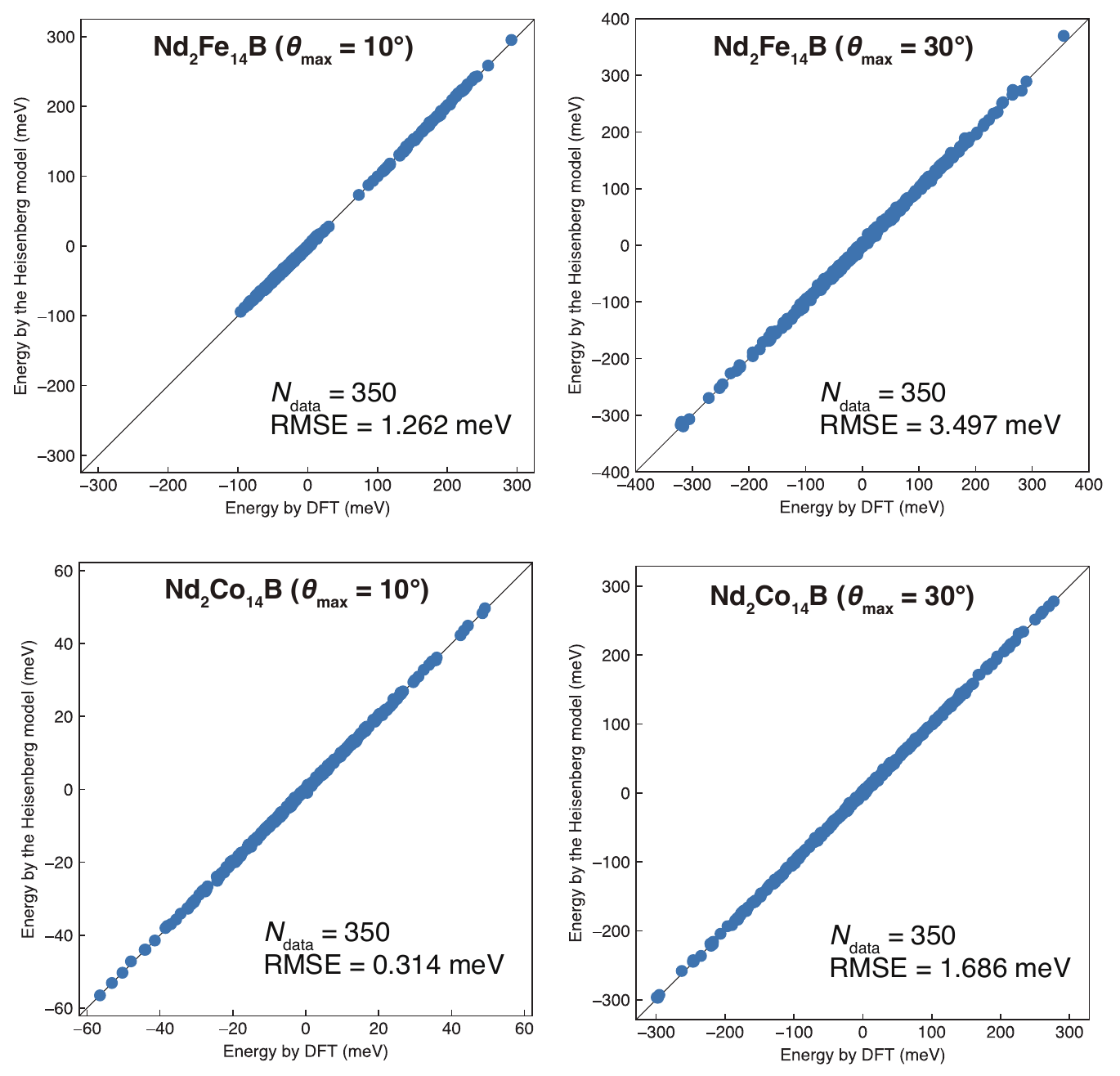}
\caption{Comparison between the energies calculated using DFT and the Heisenberg model for Nd$_2$Fe$_{14}$B and Nd$_2$Co$_{14}$B, plotted in the same manner as Fig.~\ref{energy_srmno3}.
}
\label{energy_2-14-1}
\end{figure}

\begin{figure}[h]
\centering
\includegraphics[width=\linewidth]{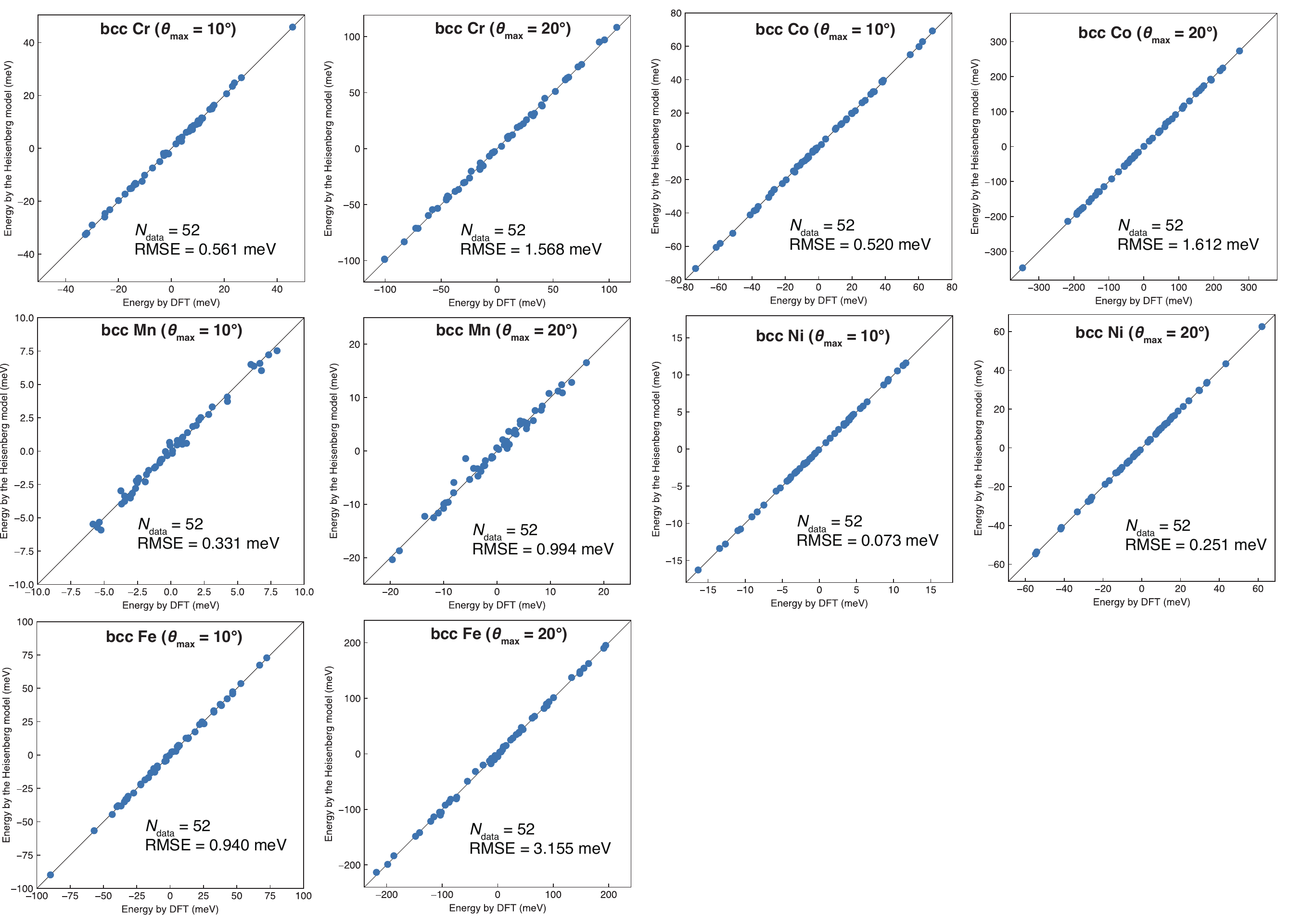}
\caption{Comparison between the energies calculated using DFT and the Heisenberg model for 3$d$ transition metals with bcc structure, plotted in the same manner as Fig.~\ref{energy_srmno3}.}
\label{energy_bcc}
\end{figure}

\begin{figure}[h]
\centering
\includegraphics[width=\linewidth]{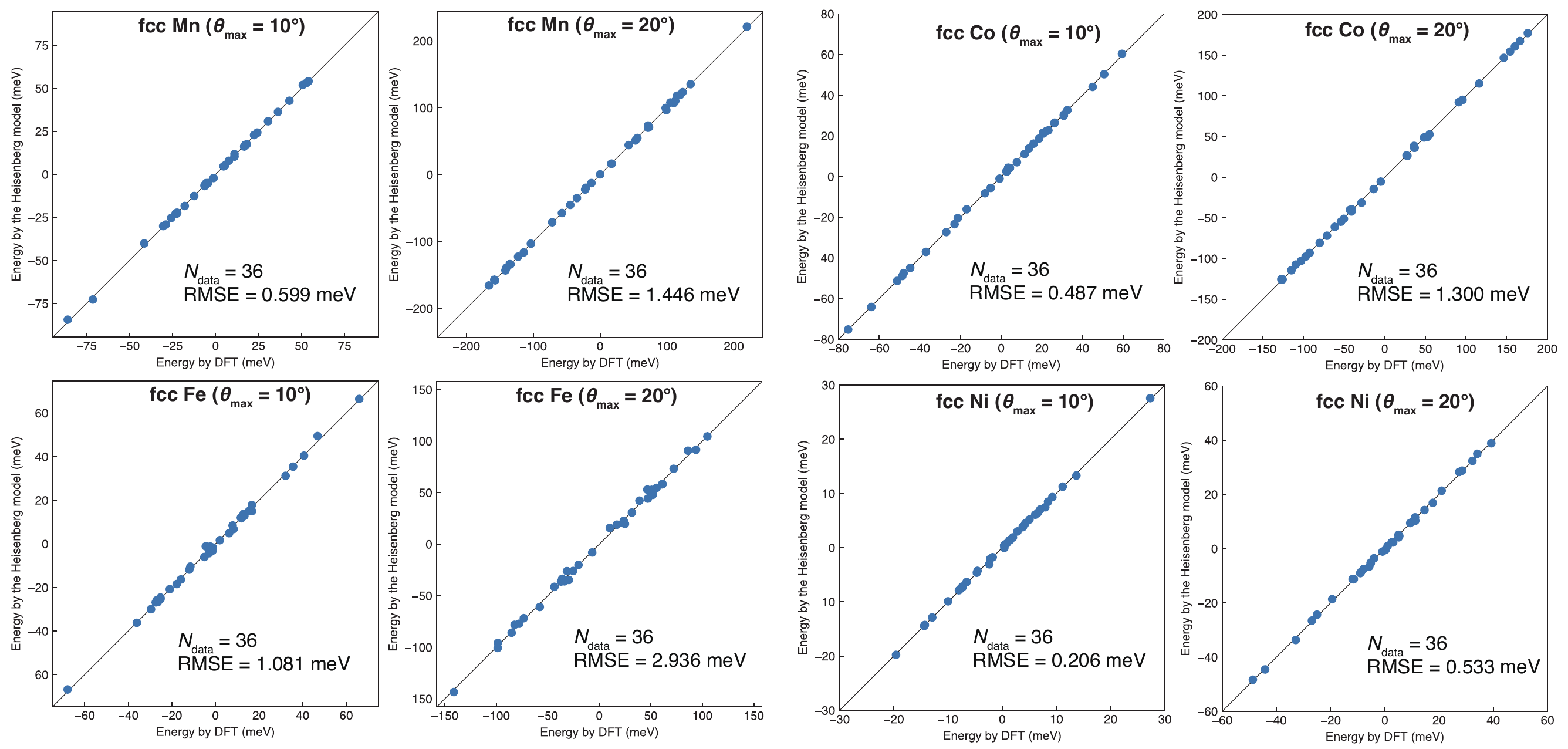}
\caption{Comparison between the energies calculated using DFT and the Heisenberg model for 3$d$ transition metals with fcc structure, plotted in the same manner as Fig.~\ref{energy_srmno3}.}
\label{energy_fcc}
\end{figure}

In this paper, we used the Heisenberg model to reproduce DFT energies.
To validate the use of this model, we present the results of regression performance of the Heisenberg model.
Figures~\ref{energy_srmno3},~\ref{energy_2-14-1},~\ref{energy_bcc}, and~\ref{energy_fcc} show the energy comparison between DFT and the Heisenberg model.
The total number of magnetic configurations $N_{\rm data}$ and the root mean squared error (RMSE) defined as
\begin{align}
{\rm RMSE} = \left[ \frac{1}{N_{\rm data}}\sum_{n=1}^{N_{\rm data}} \left(E_{\rm DFT}^{(n)} - E_{\rm model}^{(n)}\right)^2 \right]^{1/2}
\end{align}
are also presented.

\clearpage
\section{Comparison of $J_{ij}$ of bcc Fe in magnetically disordered state}
\begin{figure}[h]
\centering
\includegraphics[width=\linewidth]{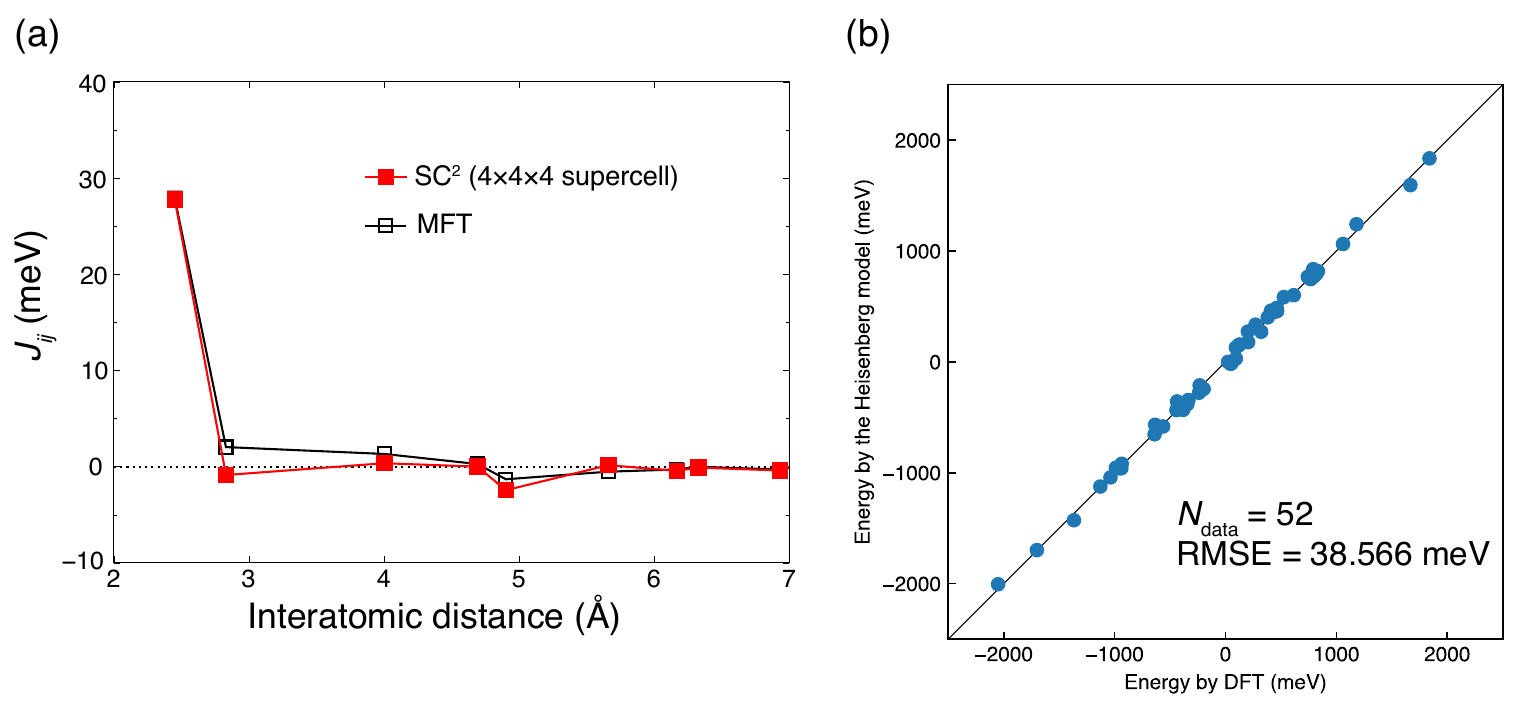}
\caption{(a) $J_{ij}$ of bcc Fe in LMD states.
The MFT-based calculations are carried out using the AkaiKKR package based on the KKR-Green’s function method.
(b) Comparison between the energies calculated using DFT and the Heisenberg model, plotted in the same manner as Fig.~\ref{energy_srmno3}.}
\label{bccfe_lmd}
\end{figure}

In the MFT method, the magnetic reference state can be chosen as a completely disordered magnetic state (referred to as the LMD state or the disordered local moment state) by using the coherent potential approximation.
In this approach, atoms with polarization directions opposite to each other form a pseudo-alloy in equal proportions. 
Figure~\ref{bccfe_lmd}(a) shows $J_{ij}$ of bcc Fe in the LMD state, calculated using both the (SC)$^2$ method ($\theta_{\rm max} = 180^\circ$) and the MFT-based method based on the KKR-Green's function method implemented in the AkaiKKR package.
As shown, the results from both methods exhibit good quantitative agreement.
Therefore, the LMD calculations in the KKR-Green's function method offer a practical and convenient choice, as they demonstrate better convergence in self-consistent field calculations compared to the (SC)$^2$ method.
We also present the energy comparison from DFT and the Heisenberg model in Fig.~\ref{bccfe_lmd}(b).

\clearpage
\section{Magnetic phase transition temperatures for 3$d$ transition metals}
\begin{figure}[h]
\centering
\includegraphics[width=80mm]{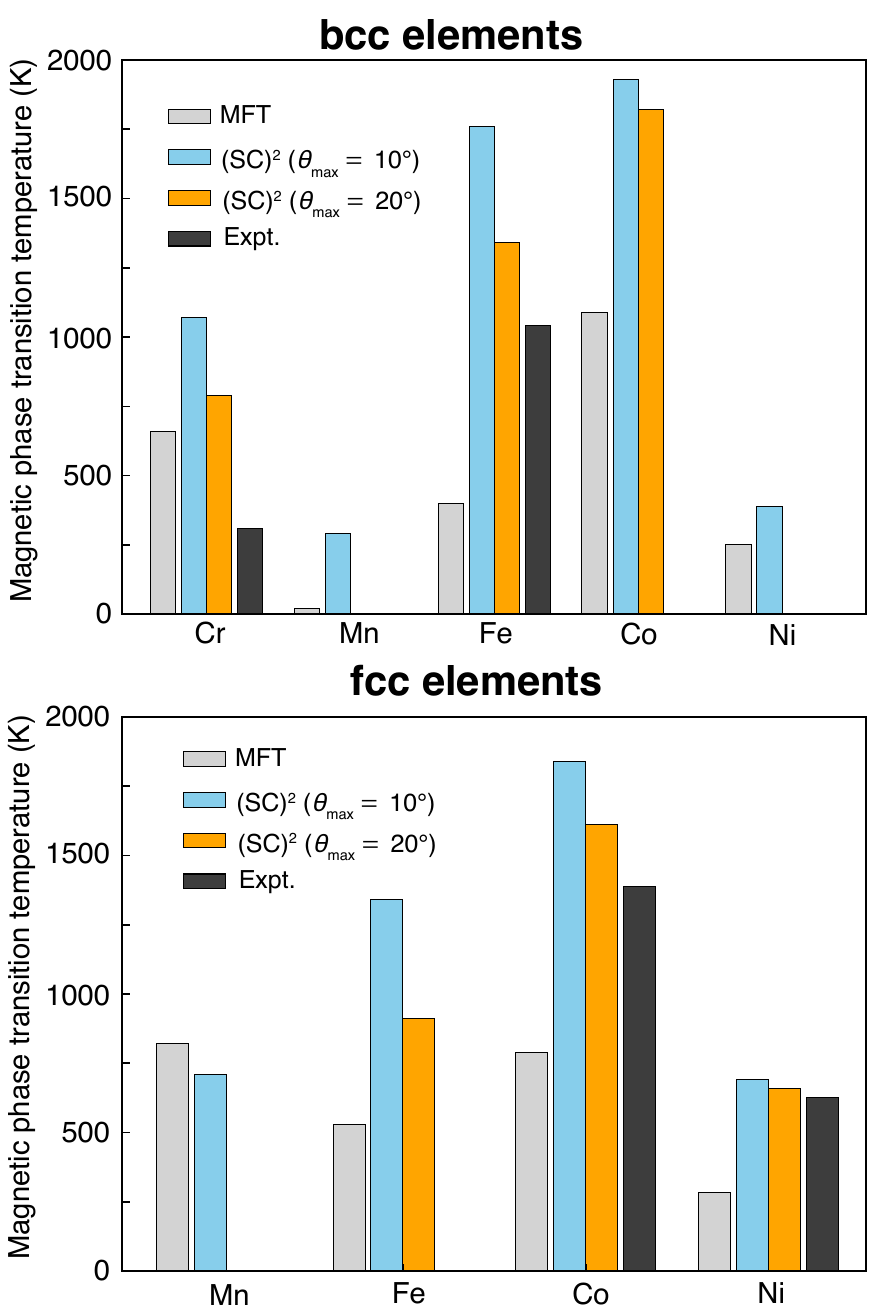}
\caption{Magnetic phase transition temperatures of elemental $3d$ metals obtained using classical Monte Carlo simulations.
Experimental values are also arranged for comparison~\cite{Haynes2016-rh}.
Note that the experimental value of bcc Cr is the one for the antiferromagnetic spin-density wave state~\cite{Fawcett1988-ax}, whereas the calculated values here are those for the type--I antiferromagnetic state.
The MFT-based results shown here are derived from our own $J_{ij}$ computed for the ground-state reference magnetic configuration; by contrast, the main text cites the results of Ref.~\cite{Shallcross2005-ka}.}
\label{tc_elements}
\end{figure}

In the main text, we report the magnetic phase transition temperatures for bcc Fe, fcc Co, and fcc Ni, which can be compared directly with experimental values.
Here, we present the results of the (SC)$^2$ method for all systems, as summarized in Fig.~\ref{tc_elements}.

\clearpage
\section{Alternative spin configuration sampling method}
In the main text, we use the uniform distribution on a spherical cap defined by the parameter $\theta_{\rm max}$.
Here we follow Ref.~\cite{Mendive-Tapia2022-wq} and give an alternative spin configuration sampling method based on the mean-field approximation (MFA) to the ferromagnetic Heisenberg model
\begin{align}
E=-\sum_{i\ne j} J_{ij}\,\hat{\bm e}_i\!\cdot\!\hat{\bm e}_j ,
\end{align}
assuming a single sublattice. Applying MFA,
\begin{align}
E \approx -\sum_i {\bm h}_i^{\rm eff}\!\cdot\!\hat{\bm e}_i, \qquad
{\bm h}_i^{\rm eff}=2\sum_j J_{ij}\,\langle\hat{\bm e}_j\rangle
=2J_0{\bm m},
\end{align}
with $J_0=\sum_j J_{ij}$ and ${\bm m}=\langle\hat{\bm e}_j\rangle$.
The self-consistent condition for ${\bm m}$ is given as 
\begin{align}
{\bm m}=\big[\coth(\beta h_i^{\rm eff})-(\beta h_i^{\rm eff})^{-1}\big]\,
\hat{\bm h}_i^{\rm eff},
\end{align}
where $h_i^{\rm eff}=|{\bm h}_i^{\rm eff}|$, $\hat{\bm h}_i^{\rm eff}={\bm h}_i^{\rm eff}/h_i^{\rm eff}$, and $\beta=1/(k_{\rm B}T)$ ($k_{\rm B}$ is the Boltzmann constant). The single-site orientational distribution is
\begin{align}
P_i(\hat{\bm e}_i)=\frac{\beta h_i^{\rm eff}}{4\pi\sinh(\beta h_i^{\rm eff})}
\exp\!\big(\beta\,{\bm h}_i^{\rm eff}\!\cdot\!\hat{\bm e}_i\big).
\end{align}
From the MFA Curie temperature
\begin{align}
T_{\rm C}^{\rm MFA}=\frac{2J_0}{3k_{\rm B}},
\end{align}
it follows that $\beta h_i^{\rm eff}=3m/\tau$ with $\tau=T/T_{\rm C}^{\rm MFA}$ and $m=|{\bm m}|$, hence
\begin{align}
P_i(\hat{\bm e}_i)=\frac{3m/\tau}{4\pi\sinh(3m/\tau)}
\exp\!\left(\frac{3\,{\bm m}\!\cdot\!\hat{\bm e}_i}{\tau}\right). \label{prob}
\end{align}
From Eq.~(\ref{prob}), temperature-dependent spin configurations can be sampled directly through the reduced temperature $\tau$, providing a simple and physically interpretable scheme.
We refer to the sampling scheme based on Eq.~(\ref{prob}) as MFA-based sampling.
Below, we apply this scheme to a $4\times4\times4$ bcc-Fe supercell and a $3\times3\times3$ fcc-Ni supercell and demonstrate that it yields results essentially identical to those obtained with the $\theta_{\rm max}$-bounded uniform distribution used in the main text.

For the direct comparison between the $\theta_{\rm max}$-bounded uniform distribution sampling and the MFA-based sampling defined by Eq.~(\ref{prob}), we match the two-spin correlation $\langle \hat{\bm e}_i \cdot \hat{\bm e}_j \rangle$ between them.
We already obtained the expression of $\langle \hat{\bm e}_i \cdot \hat{\bm e}_j \rangle$ for the uniform distribution as in Eq.~(\ref{corr_uni}).
For Eq.~(\ref{prob}), 
\begin{align}
\begin{split}
	\langle \hat{\bm e}_i \cdot \hat{\bm e}_j \rangle &= \sum_{\alpha = x, y, z} \langle e_{i,\alpha}e_{j,\alpha}\rangle \\
	&= \sum_{\alpha = x, y, z}\langle e_{i, \alpha} \rangle \langle e_{j, \alpha} \rangle \\
	&= \sum_{\alpha = x, y, z} \langle e_{\alpha}\rangle^2 \\
	&= m^2\\
	&= \left[\coth\left(3m/\tau \right)- \tau/3m \right]^2,
	\end{split}
\end{align}
where the second line uses mean-field factorization (independent spins), and the third line uses that all spins share the same single-site distribution, so that $\langle e_{i,\alpha}\rangle=\langle e_{j,\alpha}\rangle=\langle e_{\alpha}\rangle$.

\begin{figure}[h]
\centering
\includegraphics[width=\linewidth]{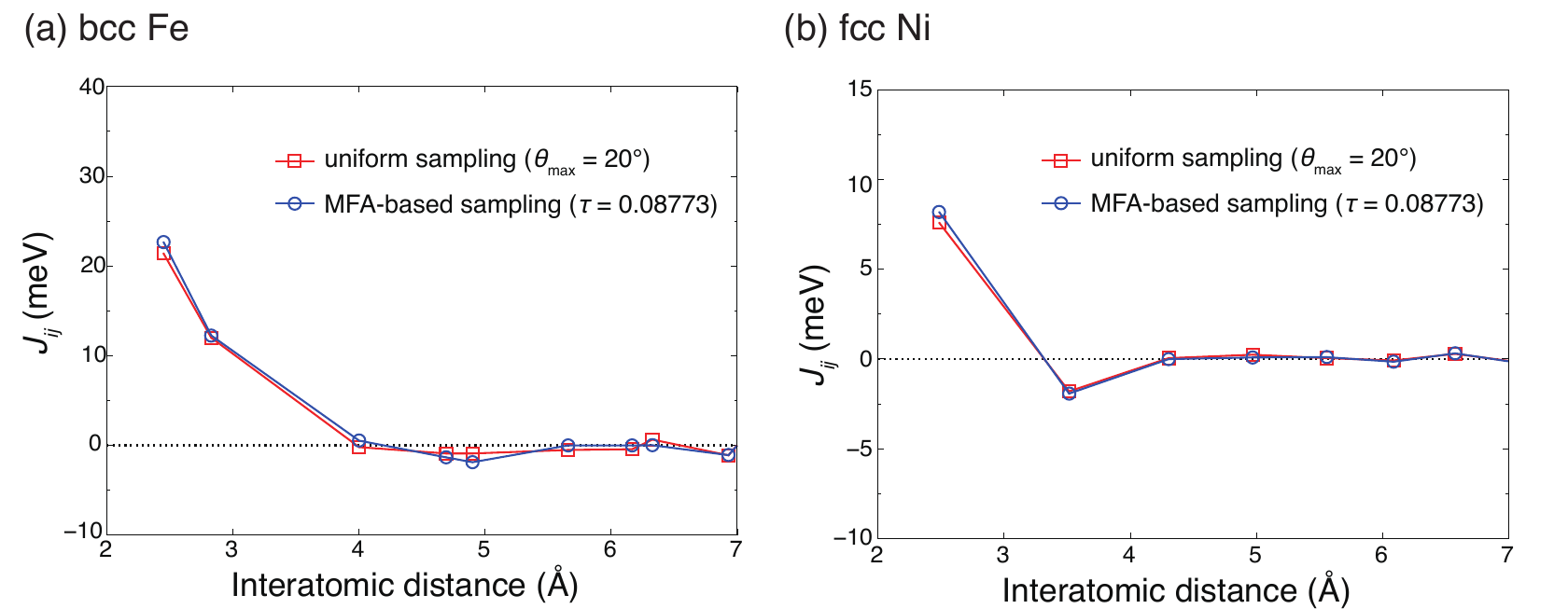}
\caption{$J_{ij}$ values obtained for (a) bcc Fe ($4\times4\times4$ supercell) and (b) fcc Ni ($3\times3\times3$ supercell) from spin configurations sampled by (i) uniform distribution sampling with $\theta_{\rm max} = 20^\circ$, and (ii) MFA-based sampling with $\tau$ chosen to match $\langle \hat{\bm e}_i \cdot \hat{\bm e}_j \rangle$ for $\theta_{\rm max} = 20^\circ$.
Note that in Figs. 6 and 7 of the main text we used $\theta_{\rm max} = 10^\circ$, whereas here we use $\theta_{\rm max} = 20^\circ$.
}
\label{sampling_comparison}
\end{figure}

Figures~\ref{sampling_comparison}(a) and (b) show the comparison of $J_{ij}$ values derived from the two sampling methods for bcc Fe and fcc Ni, respectively.
We can see that both methods yield consistent $J_{ij}$ values.
If one wishes to perform sampling in a way that is more physically intuitive, rather than relying on the parameter $\theta_{\rm max}$, one may instead choose MFA-based sampling. 
However, we emphasize again that within the scope of this study, the two are in good agreement.

In the above discussion, $T_{\rm C}^{\rm MFA}$ denotes the transition temperature within the mean-field approximation and should not be confused with the true magnetic phase transition temperature.
For $T \ge T_{\rm C}^{\rm MFA}$, not only long-range order but also short-range order completely disappears, whereas at the true transition temperature a certain degree of short-range order  remains.
The magnetic interactions that determine the magnetic phase transition temperature are themselves largely short-ranged.
In this sense, if one aims to sample magnetic configurations in the vicinity of the true magnetic phase transition, it is preferable to use the value of
$m^2 = \langle \hat{\bm e}_i \cdot \hat{\bm e}_j \rangle$
corresponding to the transition temperature obtained in previous studies, even though this procedure is somewhat empirical.
As a concrete example, in the main text we used $\tau = 0.6261$ and $0.4688$, which correspond to nearest-neighbor spin--spin correlations of $\langle \hat{\bm e}_i \cdot \hat{\bm e}_j \rangle \approx 0.5$ and $0.65$ at $T_{\rm C}$, respectively~\cite{Melnikov2019-es, Walsh2022-ci}.

\clearpage
\section{Frozen-potential decomposition of $J_{ij}$ for the elemental 3$d$ metals}
\label{frozen}
To identify the origin of the $\theta_{\rm max}$ dependence of $J_{ij}$ in the elemental 3$d$ metals, we decompose the energy of each sampled spin configuration, following Eq.~(17) of the main text, into the band-energy variation at fixed charge density and fixed magnetization magnitude, $\delta^{*}E_{\rm band}$, and a remainder that arises from the self-consistent electronic response (electron--spin coupling).
We evaluated the energy of each configuration in two ways.
The self-consistent energy $E_{\rm SCF}$ is obtained from the constrained noncollinear DFT calculation described in the main text and is the energy used for the (SC)$^2$ fits in Figs.~6--8 of the main text.
The frozen-potential energy $E_{\rm frozen}$ is obtained from the self-consistent charge and magnetization densities of the ferromagnetic reference state, in which only the direction of the magnetization density (together with the corresponding PAW one-center occupancies) is rotated rigidly within the Voronoi cell of each atom onto the sampled spin direction of that atom, while the charge density and the magnitude of the magnetization density are kept fixed.
The band energy is then evaluated non-self-consistently from this fixed density, without the constraining field.
Because the double-counting terms are invariant under this rotation, $E_{\rm frozen}$ measured from the reference state is the finite-rotation generalization of $\delta^{*}E_{\rm band}$, and the difference $E_{\rm SCF}-E_{\rm frozen}$ is the energy change due to the self-consistent relaxation of the electronic structure, which corresponds to the $\mathcal{O}(\delta n^2,\delta m^2)$ term in Eq.~(17) of the main text.
The computational conditions follow the main text ($4\times4\times4$ supercell for bcc Fe; $3\times3\times3$ supercells for fcc Co and fcc Ni).
Both calculations used the same spin configurations: 52 for bcc Fe and 36 for fcc Co and fcc Ni at each of $\theta_{\rm max}=10^\circ$ and $20^\circ$.

\begin{figure}[h]
\centering
\includegraphics[width=\linewidth]{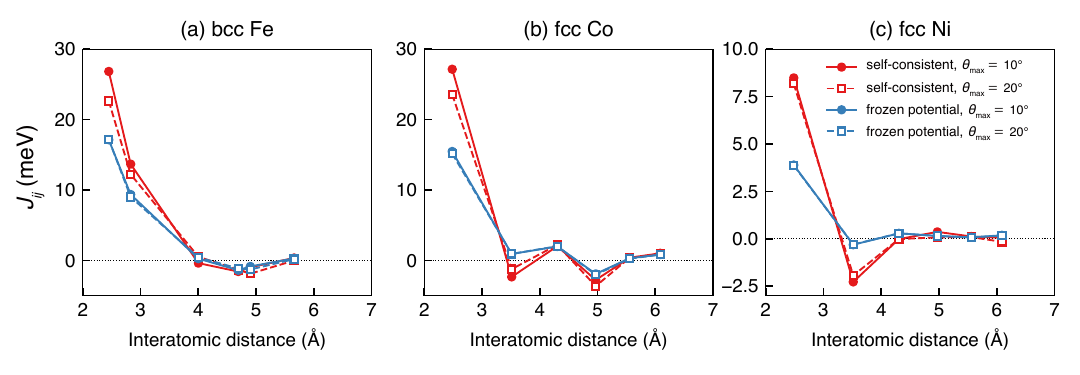}
\caption{
Heisenberg parameters $J_{ij}$ of (a) bcc Fe, (b) fcc Co, and (c) fcc Ni as a function of the interatomic distance, obtained by fitting the self-consistent energies $E_{\rm SCF}$ (red) and the frozen-potential energies $E_{\rm frozen}$ (blue) of the same sampled spin configurations, for $\theta_{\rm max}=10^\circ$ (filled circles, solid lines) and $20^\circ$ (open squares, dashed lines).
}
\label{frozen_decomposition}
\end{figure}

Figure~\ref{frozen_decomposition} compares the Heisenberg parameters obtained by fitting $E_{\rm frozen}$ and $E_{\rm SCF}$ for the same configurations.
The frozen-potential $J_{ij}$ are essentially independent of $\theta_{\rm max}$ (the sampling window): $J_{01}$ changes by $+0.3\%$ (bcc Fe), $-2.0\%$ (fcc Co), and $-0.6\%$ (fcc Ni) between $\theta_{\rm max}=10^\circ$ and $20^\circ$, whereas the self-consistent $J_{01}$ changes by $-15.7\%$, $-13.2\%$, and $-3.7\%$, respectively.
The small changes for fcc Ni in both calculations are consistent with the trend in Fig.~8 of the main text, where the $\theta_{\rm max}$ dependence of $J_{01}$ is large for bcc Fe and fcc Co and weak for fcc Ni.
The $\theta_{\rm max}$ dependence of $J_{ij}$ in these metals therefore originates predominantly from the self-consistent electronic response to the spin rotations, rather than from the nonlinearity of the fixed-density band energy.
For the 3$d$ metals, these results thus provide evidence complementary to the band-structure analysis for SrMnO$_3$ in the main text: the dependence of $J_{ij}$ on the sampled configurations is caused by electron--spin coupling, i.e., the self-consistent rearrangement of the electronic structure under finite spin rotations.

